\documentclass[twocolumn,pra,showpacs,amsmath,superscriptaddress,floatfix,longbibliography]{revtex4-1}

\usepackage{epsfig,amssymb,amsmath,amsthm,amsfonts,amsbsy,mathrsfs,amscd}
\usepackage{graphicx}
\usepackage{color}
\usepackage[colorlinks,
            linkcolor=black,
            anchorcolor=black,
            citecolor=blue]{hyperref}
\usepackage{multirow}
\usepackage{gensymb}
\usepackage{tikz}
\usepackage{mathdots}
\usepackage{yhmath}
\usepackage{booktabs}
\usepackage{bm}
\usepackage{placeins}
\usepackage{algpseudocode}
\newcounter{algorithmcounter}
\renewcommand{\thealgorithmcounter}{\arabic{algorithmcounter}}

\providecommand{\hwcode}[1]{{\ttfamily #1}}
\providecommand{\hwparam}[1]{{\ttfamily #1}}
\providecommand{\hwfile}[1]{{\ttfamily #1}}

\providecommand{\varthetahat}{\ensuremath{\hat{\vartheta}}}

\providecommand{\minC}{\ensuremath{\min|C|}}

\providecommand{\paramqzerodetune}{\hwparam{qubit0\_detune\_amplitude}}
\providecommand{\paramqonedetune}{\hwparam{qubit1\_detune\_amplitude}}

\providecommand{\paramczamp}{\hwparam{cz\_amplitude}}
\newcommand{\ii}{\mathrm{i}}
\newcommand{\Tr}{\operatorname{Tr}}
\newcommand{\Arg}{\operatorname{Arg}}
\newcommand{\ket}[1]{\left|#1\right\rangle}

\begin{document}

\title{Branch-resolved Pauli-block spectroscopy of residual conditional phase in two-qubit gates}


\author{Xudan Chai}
\email{chaixd@baqis.ac.cn}
\thanks{Corresponding author.}
\affiliation{Beijing Academy of Quantum Information Sciences, Beijing 100193, China}
\affiliation{Beijing Key Laboratory of Fault-Tolerant Quantum Computing, Beijing 100193, China}

\author{Yanwu Gu}
\affiliation{Beijing Academy of Quantum Information Sciences, Beijing 100193, China}

\author{Huiqi Xue}
\affiliation{State Key Laboratory of Low Dimensional Quantum Physics, Department of Physics, Tsinghua University, Beijing 100084, China}
\affiliation{Hefei National Laboratory, Hefei 230088, China}

\author{Kerui Li}
\affiliation{School of Medical Technology, Beijing Institute of Technology, Beijing 100081, China}

\author{Dong E. Liu}
\email{dongeliu@mail.tsinghua.edu.cn}
\thanks{Co-corresponding author.}
\affiliation{Beijing Academy of Quantum Information Sciences, Beijing 100193, China}
\affiliation{Beijing Key Laboratory of Fault-Tolerant Quantum Computing, Beijing 100193, China}

\begin{abstract}
Recent progress in quantum physics and quantum technologies is driving quantum computing from the noisy intermediate-scale (NISQ) era toward fault-tolerant and utility-scale computation.
High-precision control of two-qubit gates is among the most critical requirements in this transition, and it hinges on accurate two-qubit calibration.
For controlled-phase and CZ-style two-qubit gates, the residual conditional phase---the nonlocal \(ZZ\)-type deviation after local compensation---is weakly resolved at leading order in average infidelity and randomized benchmarking, and standard single-Pauli-sector repeated-Ramsey readout cannot reliably disentangle it from ordinary target-qubit detuning, SPAM errors, and contrast loss in long sequences.
We introduce branch-resolved Pauli-block spectroscopy to estimate the residual \(ZZ\)-rotation angle \(\vartheta_c\) accumulated per calibrated cycle, with its sign.
The protocol repeats a fixed probe for \(N\) cycles, measures the closed Pauli block \(\{IX,IY,ZX,ZY\}\), and forms branch coherences \(C_\pm\) conditioned on the control qubit; \(\vartheta_c\) splits the two branch phase slopes in opposite directions, while local target phase \(\beta_c\) shifts them together.
An echoed-cycle variant suppresses removable local terms while preserving the nonlocal contribution.
Numerical simulations with injected \(\vartheta_c\), detuning, damping, and SPAM confirm unbiased signed readout where scalar-sector alternatives fail and distinguish opposite-sign errors at equal infidelity.
On one superconducting cloud qubit--coupler pair, a single-session pulse-level calibration loop shows near-linear injected-phase response, stable branch contrast, and sign-consistent one-step point-estimate feedback.
The approach yields a low-overhead, signed per-cycle estimate of residual conditional phase that standard fidelity benchmarks underresolve at leading order.
\end{abstract}

\maketitle

\section{Introduction}
\label{sec:introduction}

Recent progress in quantum physics and quantum technologies is driving a transition from the noisy intermediate-scale (NISQ) era toward fault-tolerant and utility-scale computation~\cite{Fowler2012SurfaceCodes,Barends2014SurfaceThreshold,GoogleQEC2023SurfaceCode,GoogleWillow2025,Arute2019Supremacy,Kim2023Utility,Kandala2017VQE}.
In superconducting quantum processors~\cite{Krantz2019QuantumEngineer,Blais2021CircuitQED,Gao2021PracticalGuide}, performance requirements increasingly hinge on the fidelity and controllability of entangling operations, not on qubit number alone.\ 
High-precision implementation and calibration of two-qubit gates is therefore among the central technical challenges at this stage of development.
On present processors, a salient limitation is often a small coherent miscalibration rather than a dominant incoherent fault: such errors accumulate deterministically when gates are composed or repeated.
For controlled-phase and CZ-style gates~\cite{Rol2019CZGate,Foxen2020TwoQubitGates}, the coherent parameter most directly implicated is the residual conditional phase---equivalently, a small effective \(ZZ\) interaction after removable local phases are compensated~\cite{Sung2021Realization,Mundada2019Suppression,Zhao2021Suppression,Ni2022ScalableZZ}.
This residual governs pulse update, compilation, and repeated execution, yet is only weakly resolved at leading order by average gate infidelity and randomized benchmarking (RB)~\cite{Knill2008RB,Magesan2011RB,Magesan2012InterleavedRB}.
High-precision two-qubit calibration therefore depends on metrology that exposes such coherent residuals more directly than fidelity-based summaries alone---and, for correction, with a definite sign.

Yet realizing this metrology on hardware remains experimentally demanding.
The residual conditional phase is commonly accessed through Ramsey-type interferometry~\cite{Ramsey1950SeparatedFields,Ramsey1951PhaseShifts,Ramsey1990SeparatedFieldsReview,McKay2017VirtualZ}, as in cross-resonance tune-up and related calibration sequences~\cite{Sheldon2016CrossResonance}, but the amplified signal is typically damped, SPAM-limited, and mixed with additional target-qubit phases~\cite{Berritta2025BinarySearchRamsey,Hecht2025BeatingRamseyLimit}.
Repetition enhances sensitivity, but the same repetition amplifies decoherence~\cite{Gorini1976GKSL,Lindblad1976GKSL} and nuisance coherent terms and accumulates leakage~\cite{Wood2018Leakage,Wallman2016Leakage}; long sequences therefore deviate from an ideal single-mode fringe.
Useful metrology must therefore isolate the nonlocal phase from local common-mode contributions and remain stable under contrast loss and finite sampling in long sequences.
Existing tools address parts of this need but not the full characterization task.
RB and interleaved RB~\cite{Knill2008RB,Magesan2011RB,Magesan2012InterleavedRB} are quadratically insensitive to a small coherent phase and carry no sign; building on early symmetrized randomized characterization~\cite{Emerson2007Symmetrized}, cycle benchmarking~\cite{Erhard2019CycleBenchmarking}, character-style randomized benchmarking~\cite{Helsen2019CharacterRB}, channel-spectrum benchmarking~\cite{Gu2023CSB}, and spectral quantum tomography~\cite{Helsen2019SpectralTomography} add structured error metrics and selected channel eigenvalues, yet still target average or spectral channel information rather than a signed per-cycle estimate of the residual conditional phase.\ 
Process tomography~\cite{Poyatos1997QPT,Merkel2013SelfConsistentQPT,Bialczak2010QPT} and gate-set tomography~\cite{Nielsen2021GateSetTomography,Gu2021RLGST} can recover a coherent phase in principle, yet at a characterization cost poorly matched to routine calibration, while readouts based on a single Pauli sector become unreliable once an ordinary target detuning \(\beta_c\) mixes into the same observables.
Robust phase estimation~\cite{Kimmel2015RPE,Rudinger2017RPE,Dong2025QSPE,Gu2023NoiseResilientPE} and matrix-element amplification~\cite{Genois2024MEADD} return direct, signed coherent parameters with strong SPAM robustness, and Hamiltonian-learning methods~\cite{Burgarth2009AccessLimited,Granade2012HamiltonianLearning,Hangleiter2024HamiltonianLearning} recover effective couplings including residual \(ZZ\) terms at the model level.
Taken together, these approaches do not yet supply a signed per-cycle estimate of the residual conditional phase that remains reliable under ordinary target detuning and contrast loss in long sequences, yet is light enough for repeated-cycle metrology.

To close the gap identified above, we develop branch-resolved Pauli-block spectroscopy.
Because the target of repeated-cycle metrology is the residual conditional phase itself, the readout must return a signed per-cycle \(ZZ\) angle \(\vartheta_c\) linked to that residual by fixed conventions, rather than a fidelity-level summary alone.
Moreover, because each repetition unavoidably admixes an ordinary target-qubit phase \(\beta_c\) with the nonlocal term, lengthening the sequence improves sensitivity only at the price of contrast loss.
If \(\beta_c\) and \(\vartheta_c\) jointly drive the same target-qubit coherences, a two-moment Pauli sector cannot isolate the residual once \(\beta_c\neq 0\); the closed block \(\{IX,IY,ZX,ZY\}\), reconstructing the oscillating Pauli content under joint \(IZ\) and \(ZZ\) evolution~\cite{Patterson2019RepeatedGateTomography,Sundaresan2020HEAT}, must therefore be read through control-qubit branches, where \(\vartheta_c\) splits the branch slopes in opposite directions and \(\beta_c\) shifts them together.\ 
Branch-resolved slope readout on these trajectories then yields a signed estimate of the residual conditional phase, with an optional echoed cycle---\(XX\) refocus pulses in a dynamical-decoupling sequence~\cite{Viola1999DynamicalDecoupling,Sundaresan2020HEAT}---when removable local terms should be suppressed first.

To validate unbiased, signed branch readout of the residual conditional phase under injected phase, target detuning, damping, and SPAM, we numerically simulate the protocol of Sec.~\ref{subsec:protocol}, including an echoed-cycle variant analyzed in Sec.~\ref{subsec:numerical-validation}.
At realistic \(\beta_c\), scalar-sector readout fails on the same bit-string data whereas branch-resolved estimation recovers the injected angle and distinguishes opposite-sign coherent errors with the same average infidelity, confirming reliable signed metrology under controlled noise.
To test whether the readout transfers to the setting in which such metrology must ultimately operate, we deploy the echoed protocol on a superconducting cloud processor within a pulse-level two-qubit gate calibration closed loop on one qubit--coupler pair in a single session.
Pulse-native phase injection is near-linear, branch contrast stays high throughout the loop, and one feedback step yields a sign-consistent point-estimate suppression of the native residual conditional phase---demonstrating that branch-resolved Pauli-block spectroscopy is operationally viable on native CZ waveforms, not only under idealized noise models, while higher-statistics multi-round validation remains for future work.\ 

\section{Theory and Method}
\label{sec:theory}

In this section we develop branch-resolved Pauli-block spectroscopy of the residual conditional phase in two-qubit gates, moving from quantity conventions and the repeated-cycle model to branch readout and the experimental protocol.

\subsection{Conditional phase conventions and residual-cycle normal form}

We write two-qubit Pauli strings as tensor products, for example \(IX=I_1\otimes X_2\).
For a diagonal unitary in the computational basis,
\begin{equation}
U_{\mathrm{diag}}
=
\mathrm{diag}
\!\left(
e^{\ii\phi_{00}},
e^{\ii\phi_{01}},
e^{\ii\phi_{10}},
e^{\ii\phi_{11}}
\right),
\label{eq:udiag_general}
\end{equation}
the gauge-invariant conditional phase is
\begin{equation}
\phi_{\mathrm{cond}}
=
\phi_{00}-\phi_{01}-\phi_{10}+\phi_{11}.
\label{eq:conditional_phase_def}
\end{equation}
An ideal controlled-phase gate with target angle \(\theta_{\mathrm{tar}}\),
\begin{equation}
U_{\mathrm{CP}}(\theta_{\mathrm{tar}})
=
\mathrm{diag}(1,1,1,e^{\ii\theta_{\mathrm{tar}}}),
\label{eq:cp_gate_target}
\end{equation}
has \(\phi_{\mathrm{cond}}^{\mathrm{ideal}}=\theta_{\mathrm{tar}}\).
The controlled-phase residual reported in gate calibration is therefore
\begin{equation}
\delta_{\mathrm{CP}}
=
\phi_{\mathrm{cond}}-\theta_{\mathrm{tar}}.
\label{eq:delta_cp_def}
\end{equation}
What repeated-cycle spectroscopy makes observable is not a single-gate \(\delta_{\mathrm{CP}}\), but the coherent content of a calibrated residual cycle whose Pauli readout amplifies cycle by cycle.
The executable cycle includes designed entangling and local gates together with compile-time virtual phases; the virtual-\(Z\) offsets and echo sign structure are tracked and corrected offline into a common analysis frame before branch fitting, so each repetition contributes at leading order a net coherent rotation
\begin{equation}
U_{\mathrm{nf}}(\beta_c,\vartheta_c)
=
\exp\!\left[
-\frac{\ii}{2}
\left(
\beta_c\,IZ
+
(\omega_0+\vartheta_c)\,ZZ
\right)
\right].
\label{eq:normal_form}
\end{equation}
Here \(\beta_c\) is an ordinary target-qubit \(Z\) phase accumulated per cycle, \(\vartheta_c\) is the residual \(ZZ\)-rotation angle per cycle to be estimated, and \(\omega_0\) is a known carrier phase.
When no deliberate carrier is used, \(\omega_0=0\); if the sequence imparts a known branch-differential carrier, \(\omega_0\) is fixed by that design and subtracted in post-processing by demodulation.
A control-qubit \(ZI\) term is omitted from Eq.~\eqref{eq:normal_form} because it commutes with the readout block used below and therefore does not enter the reconstructed Pauli block \(\{IX,IY,ZX,ZY\}\).
We fix the sign convention by taking the control-branch label \(\sigma=+1\) to denote the \(Z_c=+1\) branch, i.e.\ the \(\ket{0}_c\) control state; this choice sets the signs in Eqs.~\eqref{eq:cp_vs_zz_angle} and~\eqref{eq:vartheta_hat_branchlin} and in the hardware feedback of Sec.~\ref{sec:results}, so that the estimated \(\widehat\vartheta_c\), the mapping to \(\delta_{\mathrm{CP}}\), and the applied correction all close consistently.\ 

The per-cycle \(ZZ\) parametrization in Eq.~\eqref{eq:normal_form} must be distinguished from the gate-calibration conditional-phase convention in Eq.~\eqref{eq:delta_cp_def}.
For
\begin{equation}
U_{ZZ}(\vartheta)
=
\exp\!\left(
-\frac{\ii}{2}\vartheta\,ZZ
\right),
\label{eq:uzz_def}
\end{equation}
the diagonal phases give
\begin{equation}
\phi_{\mathrm{cond}}(U_{ZZ})=-2\vartheta
\label{eq:cp_vs_zz_angle}
\end{equation}
under \(Z\ket{0}=+\ket{0}\) and \(Z\ket{1}=-\ket{1}\).
The protocol therefore returns \(\vartheta_c\) directly; a reported \(\delta_{\mathrm{CP}}\) follows from Eq.~\eqref{eq:cp_vs_zz_angle} once the hardware and compiler sign convention is fixed.

Repeated-cycle readout probes multiple depths \(N\) to amplify the small per-cycle angle; the net coherent evolution is
\begin{equation}
U_{\mathrm{nf}}^N
=
\exp\!\left[
-\frac{\ii}{2}
\left(
N\beta_c\,IZ
+
N(\omega_0+\vartheta_c)\,ZZ
\right)
\right].
\label{eq:normal_form_N}
\end{equation}
The residual \(ZZ\) contribution is therefore amplified linearly as
\begin{equation}
\Theta_N=N(\omega_0+\vartheta_c).
\label{eq:theta_N_def}
\end{equation}
Repeating the cycle lengthens the sequence and improves phase resolution, yet decoherence and SPAM erode fringe visibility at large \(N\); we therefore allow an exponential envelope \(D_N=A e^{-\Gamma N}\) on the branch amplitudes, with initial contrast \(A\) and an effective per-cycle damping rate \(\Gamma\).
With the per-cycle normal form and this repeated-cycle amplification established, we next identify which Pauli coherences rotate under the residual \(ZZ\) term and how they close into the branch readout block.

\subsection{Pauli-sector amplification of a residual \(ZZ\) rotation}
\label{sec:theory-pauli}

Write the two-qubit state in a Pauli expansion,
\begin{equation}
\rho
=
\frac{1}{4}
\sum_P r_P P,
\qquad
r_P=\Tr(P\rho),
\label{eq:pauli_expansion}
\end{equation}
where \(P\) runs over two-qubit Pauli strings.
A residual \(ZZ\) rotation generated by Eq.~\eqref{eq:uzz_def} preserves every Pauli coefficient that commutes with \(ZZ\) and rotates each anticommuting partner pair as a closed two-dimensional subspace~\cite{BlumeKohout2022PauliTaxonomy}.
If \(\{P,Q\}=0\) and both anticommute with \(ZZ\), then \(P\), \(Q\), and \(ZZ\) span a three-generator algebra and the coefficients \((r_P,r_Q)\) undergo a planar rotation under \(U_{ZZ}(\vartheta)\).
Table~\ref{tab:closed_sectors} lists the four single-coherence sectors of this type in a control--target labeling with qubit~1 as control and qubit~2 as target; the partner string is the Pauli coefficient directly populated by the \(ZZ\) dynamics from an initial probe on the first column.

\begin{table}[t]
\caption{Closed single-coherence Pauli sectors under \(U_{ZZ}(\vartheta)=\exp[-\ii\vartheta ZZ/2]\).
Each row is an equivalent two-dimensional readout plane: an initial probe coherence on the left is rotated into its partner coefficient on the right.
The signs follow Eq.~\eqref{eq:IX_ZY_rotation} with an initial probe coefficient \(+1\).
Swapping the control and target qubits relabels the rows but leaves the physics unchanged.}
\label{tab:closed_sectors}
\centering
\begin{ruledtabular}
\begin{tabular}{ccc}
Probe & Partner & Partner at small \(\vartheta\) \\
\hline
\(IX\) & \(ZY\) & \(r'_{ZY}=+\sin\vartheta\) \\
\(IY\) & \(ZX\) & \(r'_{ZX}=-\sin\vartheta\) \\
\(XI\) & \(YZ\) & \(r'_{YZ}=+\sin\vartheta\) \\
\(YI\) & \(XZ\) & \(r'_{XZ}=-\sin\vartheta\) \\
\end{tabular}
\end{ruledtabular}
\end{table}

\begin{figure}[t]
    \centering
    \includegraphics[width=0.98\columnwidth]{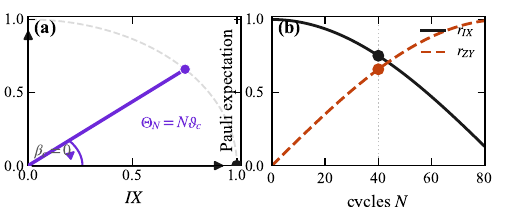}
    \caption{
    Two-dimensional Ramsey response in the \(\{IX,ZY\}\) sector for probe \(\ket{+}\ket{+}\) when \(\beta_c=0\) and \(\omega_0=0\).\ 
    (a)~Geometric rotation of the quadrature \(q_N=r_{IX}+\ii r_{ZY}\) by \(\Theta_N=N\vartheta_c\) [Eq.~\eqref{eq:IX_ZY_rotation}].
    (b)~Pauli expectations \(r_{IX}(N)=\cos(N\vartheta_c)\) and \(r_{ZY}(N)=\sin(N\vartheta_c)\); partner moments \(r_{IY}\) and \(r_{ZX}\) remain zero.
    The vertical guide marks the cycle number used in panel~(a).
    }
    \label{fig:IX_ZY_sector}
\end{figure}

Among the four equivalent rows of Table~\ref{tab:closed_sectors}, we adopt the first under the branch-resolved probe \(\ket{+}_c\ket{+}_t\): preparing the target qubit in \(\ket{+}\) seeds an \(IX\) coherence, which the residual \(ZZ\) rotation drives into its \(ZY\) partner.
In that sector the dynamics take the explicit form
\begin{equation}
\begin{pmatrix}
r'_{IX}\\
r'_{ZY}
\end{pmatrix}
=
\begin{pmatrix}
\cos\vartheta & -\sin\vartheta\\
\sin\vartheta & \cos\vartheta
\end{pmatrix}
\begin{pmatrix}
r_{IX}\\
r_{ZY}
\end{pmatrix}.
\label{eq:IX_ZY_rotation}
\end{equation}
When \(\beta_c=0\) and \(\omega_0=0\), Eqs.~\eqref{eq:normal_form_N} and \eqref{eq:IX_ZY_rotation} reduce to a two-dimensional Ramsey response,
\(r_{IX}(N)=\cos(N\vartheta_c)\) and \(r_{ZY}(N)=\sin(N\vartheta_c)\), so the complex quadrature \(q_N=r_{IX}(N)+\ii r_{ZY}(N)=\exp(\ii N\vartheta_c)\) encodes the amplified residual angle \(\Theta_N=N\vartheta_c\).
Figure~\ref{fig:IX_ZY_sector} shows this rotation in the \(\{IX,ZY\}\) plane.
Resolving the two control-qubit branches, however, requires initial population on both \(Z_c=\pm 1\) branches: the control qubit must be prepared off the \(Z\) axis so that the differential estimator of Eq.~\eqref{eq:vartheta_hat_branchlin} has two slopes to compare.
For branch-resolved readout we use a control-balanced target-coherence probe, taken here as \(\ket{+}_c\ket{+}_t\).
In the \(\{IX,IY,ZX,ZY\}\) block this gives \(r_{IX}(0)=1\) and \(r_{IY}(0)=r_{ZX}(0)=r_{ZY}(0)=0\), with \(r_{ZI}(0)=0\).
The additional control-coherence components of the pure \(\ket{+}_c\) preparation evolve in Pauli sectors outside this block under Eq.~\eqref{eq:normal_form} and do not enter the branch reconstruction.\ 
A control qubit prepared in a \(Z\) eigenstate such as \(\ket{0}\) would instead populate only a single branch (\(C_-\equiv 0\)) and leave \(r_{ZI}=1\), so the branch-differential angle \(\widehat\vartheta_c=(s_+-s_-)/2\) could not be formed.
Numerical validations and the cloud experiment reported in Sec.~\ref{sec:results} use this \(\ket{+}\ket{+}\) probe throughout, keeping the branch readout and estimator unchanged.

The \(\{IX,ZY\}\) plane alone is not sufficient for the general normal form in Eq.~\eqref{eq:normal_form}.
The ordinary target-qubit phase \(\beta_c IZ\) commutes with \(ZZ\) but rotates the target \(X\)--\(Y\) coherences, mixing \(IX\) with \(IY\) and \(ZX\) with \(ZY\).
Starting from an \(IX\) probe, joint evolution under \(IZ\) and \(ZZ\) therefore closes the four-dimensional block \(\{IX,IY,ZX,ZY\}\) rather than a single two-dimensional sector, so \(\beta_c\) and \(\vartheta_c\) must be separated after branch resolution.

Define
\begin{equation}
B_N=N\beta_c,
\qquad
\Theta_N=N(\omega_0+\vartheta_c).
\label{eq:BN_ThetaN}
\end{equation}
Because \([IZ,ZZ]=0\), the per-cycle \(IZ\) and \(ZZ\) rotations commute and their net effect factorizes; in the ideal limit, composing the \(IZ\) evolution with the \(ZZ\) partner rotations of Table~\ref{tab:closed_sectors} yields
\begin{align}
r_{IX}(N)&=D_N\cos B_N\cos\Theta_N,
\label{eq:rIX_full}\\
r_{IY}(N)&=D_N\sin B_N\cos\Theta_N,
\label{eq:rIY_full}\\
r_{ZX}(N)&=-D_N\sin B_N\sin\Theta_N,
\label{eq:rZX_full}\\
r_{ZY}(N)&=D_N\cos B_N\sin\Theta_N.
\label{eq:rZY_full}
\end{align}
The scalar angle \(\Phi_N^{\mathrm{sec}}=\operatorname{atan2}[r_{ZY}(N),r_{IX}(N)]\) tracks the \(\{IX,ZY\}\) sector of Table~\ref{tab:closed_sectors} and is useful for visualization, but it is not a robust estimator when \(\beta_c\neq 0\), because the \(IY\) and \(ZX\) components neglected by a two-moment readout enter through \(\cos B_N\) and induce spurious jumps unrelated to \(\vartheta_c\).

The branch-resolved construction separates the common-mode and differential phases.
Define
\begin{align}
u_N&=r_{IX}(N)+\ii r_{IY}(N),
\label{eq:u_def}\\
v_N&=r_{ZX}(N)+\ii r_{ZY}(N),
\label{eq:v_def}
\end{align}
Operationally, branch-resolved coordinates are assembled from the measured Pauli block via the unnormalized branch numerators
\begin{equation}
C_\sigma^{\mathrm{num}}(N)=u_N+\sigma v_N,
\qquad
\sigma=\pm 1.
\label{eq:C_branch_num}
\end{equation}
The normalized conditional branch coherences are
\begin{equation}
\begin{aligned}
C_\sigma(N)
&=
\frac{C_\sigma^{\mathrm{num}}(N)}{1+\sigma r_{ZI}(N)}\\
&=
\frac{r_{IX}(N)+\ii r_{IY}(N)+\sigma\bigl[r_{ZX}(N)+\ii r_{ZY}(N)\bigr]}{1+\sigma r_{ZI}(N)}.
\end{aligned}
\label{eq:C_branch}
\end{equation}
For the balanced probe used below, \(r_{ZI}(N)\simeq 0\), so \(C_\sigma\simeq C_\sigma^{\mathrm{num}}\); the protocol below uses \(\ket{+}\ket{+}\) precisely to maintain this regime.\ 
When \(r_{ZI}\) is not negligible, readout should employ the normalized \(C_\sigma\) of Eq.~\eqref{eq:C_branch} rather than \(C_\sigma^{\mathrm{num}}\) alone.
In both numerical processing and hardware analysis we apply Eq.~\eqref{eq:C_branch} with \(r_{ZI}(N)\) extracted from the same joint readout used for the Pauli block [Algorithm~\ref{alg:protocol}]; for \(\ket{+}\ket{+}\) this reduces to the numerator alone.\ 
Using Eqs.~\eqref{eq:rIX_full}--\eqref{eq:rZY_full} with \(r_{ZI}(N)\simeq 0\),
\begin{equation}
C_\sigma(N)
\simeq
C_\sigma^{\mathrm{num}}(N)
=
D_N
\exp\!\left[
\ii N\left(\beta_c+\sigma\omega_0+\sigma\vartheta_c\right)
\right]
\label{eq:C_ideal}
\end{equation}
in the ideal coherent limit.
The ordinary target-qubit phase \(\beta_c\) is therefore common mode in the branch phases, whereas \(\vartheta_c\) appears with opposite sign on the two branches.
Carrier demodulation gives
\begin{equation}
\widetilde{C}_\sigma(N)
=
\exp(-\ii\sigma\omega_0 N)\,C_\sigma(N)
\simeq
D_N
\exp\!\left[
\ii N\left(\beta_c+\sigma\vartheta_c\right)
\right],
\label{eq:C_demod}
\end{equation}
which is the input sequence for the branch-resolved readout below.

\begin{figure}[t]
    \centering
    \includegraphics[width=\columnwidth]{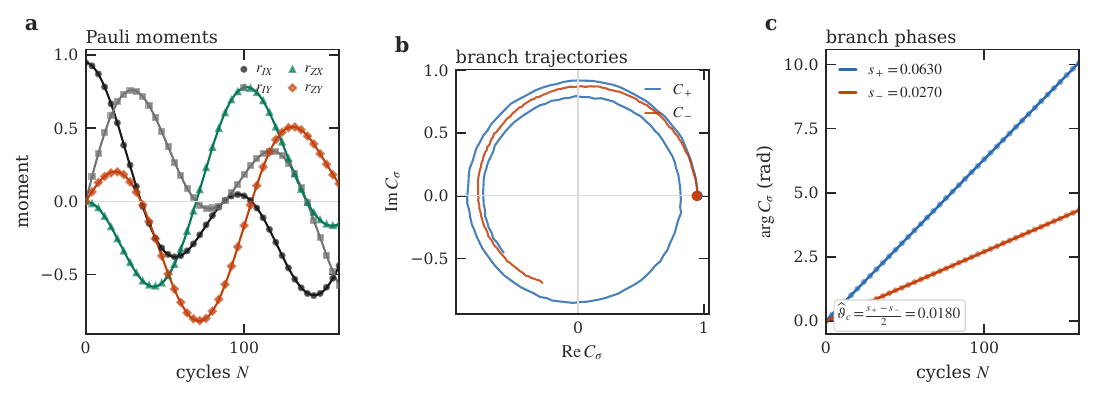}
    \caption{
    Pauli-block closure and branch-resolved readout under a residual \(ZZ\) rotation (simulated example; probe \(\ket{+}\ket{+}\)).\ 
    (a)~Pauli moments \(\{r_{IX},r_{IY},r_{ZX},r_{ZY}\}\) versus cycle number \(N\), compared with Eqs.~\eqref{eq:rIX_full}--\eqref{eq:rZY_full} including contrast decay.
    (b)~Branch coherences \(C_\pm\) [Eq.~\eqref{eq:C_branch}; numerators \(C_\pm^{\mathrm{num}}=u_N\pm v_N\) when \(r_{ZI}\simeq 0\)] in the complex plane.
    (c)~Unwrapped branch phases versus \(N\); linear fits give slopes \(s_\pm\) and \(\widehat\vartheta_c=(s_+-s_-)/2\) [Eq.~\eqref{eq:vartheta_hat_branchlin}].
    The target-qubit phase \(\beta_c\) is common mode in \(s_\pm\), whereas \(\vartheta_c\) is the differential half-splitting.
    }
    \label{fig:block_closure}
\end{figure}

Given demodulated branch sequences \(\widetilde{C}_\sigma(N_m)\) at cycle numbers \(\{N_m\}\), we estimate \(\vartheta_c\) by branch-resolved linear Ramsey readout.
For each branch \(\sigma\), form the unwrapped phase
\begin{equation}
\phi_\sigma(N_m)
=
\mathrm{unwrap}\,
\Arg\widetilde{C}_\sigma(N_m)
\label{eq:unwrap_phase}
\end{equation}
and fit a weighted linear slope
\begin{equation}
s_\sigma
=
\arg\min_s
\sum_m
w_{\sigma,m}
\left[
\phi_\sigma(N_m)-sN_m-b_\sigma
\right]^2,
\label{eq:branch_slope}
\end{equation}
with weights \(w_{\sigma,m}\propto|\widetilde{C}_\sigma(N_m)|\).
Equations~\eqref{eq:C_demod}--\eqref{eq:branch_slope} imply \(s_+\approx\beta_c+\vartheta_c\) and \(s_-\approx\beta_c-\vartheta_c\), so the per-cycle estimate is
\begin{equation}
\widehat\vartheta_c
=
\frac{s_+-s_-}{2}.
\label{eq:vartheta_hat_branchlin}
\end{equation}
Equation~\eqref{eq:vartheta_hat_branchlin} is the analytic branch-linear estimator: it is direct, matches the linear phase structure of Eqs.~\eqref{eq:C_ideal}--\eqref{eq:C_demod}, and makes the differential-slope interpretation of \(\vartheta_c\) explicit.
In the presence of shot noise, per-cycle damping, and SPAM, however, we adopt the equivalent matrix-pencil pole readout of Appendix~\ref{app:readout-comparison} [Algorithm~\ref{alg:pencil}]---which analyzes the demodulated branch sequences as damped discrete exponentials---as the working estimator used in all numerical and hardware results below.\ 
The two readouts coincide within statistical scatter in the ideal, linear-phase limit, as shown by the estimator-comparison tests of Appendix~\ref{app:readout-comparison} (Fig.~\ref{fig:app_readout_comparison}); Eq.~\eqref{eq:vartheta_hat_branchlin} thus serves as the transparent analytic limit of the matrix-pencil estimate.
Figure~\ref{fig:block_closure} illustrates the full chain under \(\beta_c\neq 0\): four-moment reconstruction, branch coherences \(C_\pm\), and differential phase extraction.
The relations above are derived in this leading-order coherent model, with contrast loss captured by \(D_N\) and \(r_{ZI}(N)\simeq 0\) so that \(C_\sigma\simeq C_\sigma^{\mathrm{num}}\); scope and assumptions are collected in Sec.~\ref{par:scope-limits} below.

\subsection{Protocol}
\label{subsec:protocol}

\begin{figure}[t]
    \centering
    \includegraphics[width=\columnwidth]{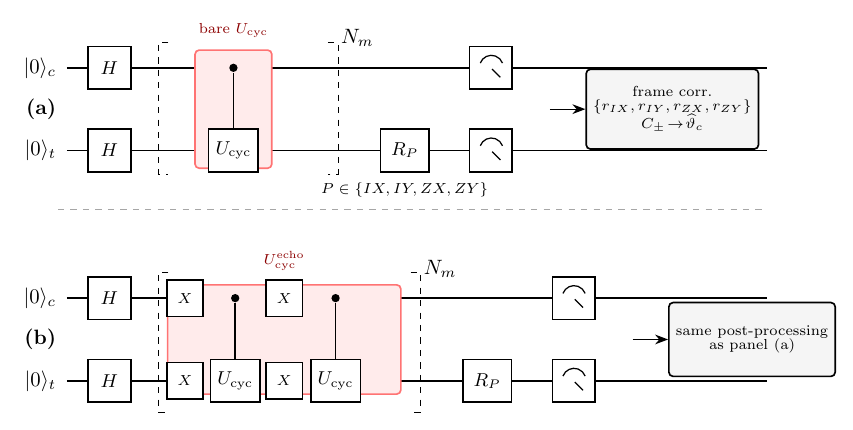}
    \caption{
    Protocol circuit for branch-resolved Pauli-block spectroscopy of a residual \(ZZ\) rotation.
    (a)~Bare residual cycle: the probe state \(\ket{+}\ket{+}\) [Eq.~\eqref{eq:probe_state}] is prepared by applying \(H\) to each qubit initialized in \(\ket{0}\), then evolved by \(N_m\) repetitions of \(U_{\mathrm{cyc}}\) [Eq.~\eqref{eq:total_evolution}].
    This template is used in the numerical validations of Figs.~\ref{fig:scalar_vs_branch}--\ref{fig:noise_robustness_branch} and \ref{fig:complementarity}.
    (b)~Echoed supercycle [Eq.~\eqref{eq:echo_cycle}]: each repeated unit is \(U_{\mathrm{cyc}}^{\mathrm{echo}}=(XX)\,U_{\mathrm{cyc}}\,(XX)\,U_{\mathrm{cyc}}\), with simultaneous \(X\) refocus pulses on both qubits; the same readout and post-processing follow.
    The hardware experiment (Sec.~\ref{subsec:hardware-results}) implements this echoed template literally on native \hwcode{CZ} pulses [Eq.~\eqref{eq:echo-cycle-hw}].
    Two target-qubit pre-rotations \(R_X,R_Y\) map the fixed-frame Pauli block \(\{IX,IY,ZX,ZY\}\) onto joint computational readout (the \(X\)-basis setting decodes \(IX,ZX\); the \(Y\)-basis setting decodes \(IY,ZY\)) [Eq.~\eqref{eq:pauli_from_counts}].\ 
    Offline frame correction, branch reconstruction, and differential phase fitting yield \(\widehat\vartheta_c\) [Eq.~\eqref{eq:vartheta_hat_branchlin}].
    }
    \label{fig:protocol_circuit}
\end{figure}

The protocol follows the usual calibration layout: state preparation, application of a repeatable residual-cycle gate set, Pauli-tomographic measurement, and classical post-processing with the branch-resolved estimator of Sec.~\ref{sec:theory-pauli}, in the spirit of repeated-cycle Pauli Hamiltonian tomography~\cite{Patterson2019RepeatedGateTomography,Sundaresan2020HEAT}.
In the present experiment this realizes repeated-cycle amplification of the per-cycle residual \(ZZ\) angle \(\vartheta_c\) in the normal form of Eq.~\eqref{eq:normal_form}; Figure~\ref{fig:protocol_circuit} and Algorithm~\ref{alg:protocol} outline the circuit and end-to-end flow from shot counts to \(\widehat\vartheta_c\), with each stage specified below.

\paragraph{State preparation (probe).}
Qubit~1 is the control and qubit~2 is the target.
We prepare the product probe
\begin{equation}
\ket{\psi_0}=\ket{+}\ket{+},
\label{eq:probe_state}
\end{equation}
which seeds the \(IX\) coherence of the first row of Table~\ref{tab:closed_sectors}.
Placing the control qubit on the equator (\(r_{ZI}=0\)) populates both \(Z_c=\pm 1\) branches equally, so the differential branch readout of Eq.~\eqref{eq:vartheta_hat_branchlin} has two slopes to compare; a control prepared in a \(Z\) eigenstate such as \(\ket{0}\) would leave only one branch (\(C_-\equiv 0\)) and cannot resolve \(\vartheta_c\).
In the ideal limit the oscillating block \(\{IX,IY,ZX,ZY\}\) then satisfies \(r_{IX}(0)=1\) and \(r_{IY}(0)=r_{ZX}(0)=r_{ZY}(0)=0\) [Eqs.~\eqref{eq:rIX_full}--\eqref{eq:rZY_full} at \(N=0\)], so \(r_{ZI}(N)\simeq 0\) and \(C_\sigma\simeq C_\sigma^{\mathrm{num}}\) [Eqs.~\eqref{eq:C_branch_num}--\eqref{eq:C_branch}].
Any other Table~\ref{tab:closed_sectors} row may be substituted provided the control remains off the \(Z\) axis and the same analysis frame is maintained throughout; Eq.~\eqref{eq:probe_state} fixes the convention for the circuit diagram and algorithm below.

\paragraph{Residual-cycle evolution.}
For each sampled cycle number \(N_m\) in a preset list \(\{N_m\}_{m=1}^{M}\), apply
\begin{equation}
U_{\mathrm{tot}}(N_m)=U_{\mathrm{cyc}}^{N_m},
\label{eq:total_evolution}
\end{equation}
where \(U_{\mathrm{cyc}}\) is the calibrated residual cycle whose coherent part is modeled by Eq.~\eqref{eq:normal_form}.
Repeated application amplifies the per-cycle angle as in Eq.~\eqref{eq:normal_form_N}; the probe evolves within the fixed-frame block of Eqs.~\eqref{eq:rIX_full}--\eqref{eq:rZY_full}, which the measurement below samples at each \(N_m\).
The list \(\{N_m\}\) is chosen to span a well-conditioned accumulated phase range; for injected calibration at small \(|\vartheta_c|\), the maximum index may be rescaled adaptively while keeping the readout formulas fixed.
When removable per-cycle \(IZ\) accumulation must be suppressed before readout, \(U_{\mathrm{cyc}}\) may be replaced by an echoed template analyzed in Sec.~\ref{subsec:numerical-validation} [Eq.~\eqref{eq:echo_cycle}]; in that case the repeated unit indexed by \(N\) is one echoed cycle, and \(\widehat\vartheta_c\) is reported per echoed cycle (see Sec.~\ref{subsec:numerical-validation}).\ 
When the echoed template contains two native entangling blocks, the angle returned by the estimator is the residual angle per echoed supercycle; conversion to a per-native-\hwcode{CZ} residual therefore requires dividing by the number of native \hwcode{CZ} blocks in the cycle, unless stated otherwise.\ 

\paragraph{Measurement.}
After \(U_{\mathrm{tot}}(N_m)\), read out the four fixed-frame Pauli moments of the oscillating block \(\{IX,IY,ZX,ZY\}\) established in state preparation.
These four moments require only \emph{two} target-qubit pre-rotation settings rather than four independent measurement bases: a target \(X\)-basis setting (\(R=Y_{-\pi/2}\)) yields \(IX\) and \(ZX\) simultaneously from the target-marginal and control--target-correlated decodings of the same joint readout, while a target \(Y\)-basis setting (\(R=X_{\pi/2}\)) yields \(IY\) and \(ZY\).
In each setting the target pre-rotation maps the chosen coherence onto the computational readout axis, followed by joint computational-basis measurement on both qubits.\ 
Repeating the circuit for each basis setting \(R_B\) (\(B\in\{X,Y\}\)) and accumulating shots yields empirical estimates
\begin{equation}
\widehat r_P(N_m)
=
\frac{N_P(+\,|\,m)-N_P(-\,|\,m)}{N_P(+\,|\,m)+N_P(-\,|\,m)},
\label{eq:pauli_from_counts}
\end{equation}
where \(P\in\{IX,IY,ZX,ZY\}\) and \(N_P(\pm\,|\,m)\) are the shot counts associated with the \(\pm1\) eigenvalue of Pauli string \(P\) under the chosen pre-rotation and readout decoding.
In practice, therefore, the four moments are obtained from the two target \(X/Y\) pre-rotation settings above with shared joint readout, each setting supplying one \(I\)-type and one \(Z\)-type moment; the mapping from raw bit strings to \(\widehat r_P\) is fixed by the compiler and readout assignment and is applied identically at every \(N_m\).\ 
These four estimates supply the phase-versus-\(N_m\) data on which branch coherences, and hence \(\widehat\vartheta_c\), are formed in the next stage.

\paragraph{Classical post-processing.}
From the Pauli moments \(\widehat r_P(N_m)\) measured above, apply the branch-resolved estimator of Sec.~\ref{sec:theory-pauli}.
All deterministic frame updates---virtual-\(Z\) phases applied during compilation, echo-induced sign flips, toggling-frame conventions, and any recorded random Pauli twirl labels---are removed \emph{before} averaging so that every \(\widehat r_P(N_m)\) refers to the same fixed analysis frame as Eqs.~\eqref{eq:rIX_full}--\eqref{eq:rZY_full}.
Form the complex quadratures
\begin{equation}
\begin{aligned}
\widehat u_m&=\widehat r_{IX}(N_m)+\ii\widehat r_{IY}(N_m),\\
\widehat v_m&=\widehat r_{ZX}(N_m)+\ii\widehat r_{ZY}(N_m),
\end{aligned}
\label{eq:uhat_vhat}
\end{equation}
as in Eqs.~\eqref{eq:u_def}--\eqref{eq:v_def}, form branch numerators \(\widehat C_\sigma^{\mathrm{num}}(N_m)=\widehat u_m+\sigma\widehat v_m\) [Eq.~\eqref{eq:C_branch_num}], estimate \(\widehat r_{ZI}(N_m)\) from the same joint readout, and form normalized branch coherences \(\widehat C_\sigma(N_m)=\widehat C_\sigma^{\mathrm{num}}(N_m)/[1+\sigma\widehat r_{ZI}(N_m)]\) [Eq.~\eqref{eq:C_branch}].
Under the \(\ket{+}\ket{+}\) probe, \(r_{ZI}(N)\simeq 0\) and \(\widehat C_\sigma\simeq\widehat C_\sigma^{\mathrm{num}}\); demodulate to \(\widetilde{C}_\sigma(N_m)\) via Eq.~\eqref{eq:C_demod}.
Unwrap the branch phases [Eq.~\eqref{eq:unwrap_phase}], fit the weighted slopes \(s_\sigma\) in Eq.~\eqref{eq:branch_slope}, and obtain \(\widehat\vartheta_c=(s_+-s_-)/2\) from Eq.~\eqref{eq:vartheta_hat_branchlin}.
In all reported numerical and hardware results, this final slope step is carried out with the matrix-pencil pole estimator of Algorithm~\ref{alg:pencil}, of which Eq.~\eqref{eq:vartheta_hat_branchlin} is the transparent linear-phase limit.
When a cycle duration \(T_c\) is defined, the corresponding residual coupling rate is reported as \(\widehat\zeta_{\mathrm{eff}}=2\widehat\vartheta_c/T_c\); a controlled-phase residual follows from Eq.~\eqref{eq:cp_vs_zz_angle} once the hardware sign convention is fixed.
Algorithm~\ref{alg:protocol} collects the four protocol stages above in pseudocode; Sec.~\ref{sec:results} validates whether the resulting \(\widehat\vartheta_c\) remains unbiased and sign-resolving under noise and on hardware.

\begin{figure}[t]
\refstepcounter{algorithmcounter}\label{alg:protocol}
\noindent\textbf{Algorithm~\thealgorithmcounter.} Branch-resolved Pauli-block protocol for estimating \(\widehat\vartheta_c\)
\vspace{0.35em}
\begin{algorithmic}[1]
\Require Calibrated residual cycle \(U_{\mathrm{cyc}}\); sampled cycle numbers \(\{N_m\}\); known carrier \(\omega_0\) (if any); target pre-rotations \(R_X,R_Y\) selecting the \(X\)- and \(Y\)-basis readout
\Ensure Per-cycle residual \(ZZ\) angle \(\widehat\vartheta_c\)
\For{each cycle number \(N_m \in \{N_m\}\)}
  \For{each target basis \(B\in\{X,Y\}\)}\ 
    \For{each shot in the setting ensemble}
      \State \textbf{Prepare} a fresh \(\ket{\psi_0}=\ket{+}\ket{+}\) \Comment{Eq.~\eqref{eq:probe_state}; re-prepared every circuit/shot}\ 
      \State \textbf{Evolve} \(U_{\mathrm{tot}}(N_m)=U_{\mathrm{cyc}}^{N_m}\) \Comment{Eq.~\eqref{eq:total_evolution}}
      \State \textbf{Measure} apply \(R_B\); read out both qubits in the computational basis
    \EndFor
    \State Decode joint bit strings: \(B{=}X\!\to\!\widehat r_{IX},\widehat r_{ZX}\); \(B{=}Y\!\to\!\widehat r_{IY},\widehat r_{ZY}\); each basis also contributes \(\widehat r_{ZI}\) \Comment{Eq.~\eqref{eq:pauli_from_counts}}
  \EndFor
  \State Correct all \(\widehat r_P(N_m)\) into the fixed analysis frame; set \(\widehat r_{ZI}(N_m)\) as the mean control-\(Z\) expectation over the two bases
  \State \(\widehat u_m \gets \widehat r_{IX}(N_m)+\ii\widehat r_{IY}(N_m)\), \quad \(\widehat v_m \gets \widehat r_{ZX}(N_m)+\ii\widehat r_{ZY}(N_m)\) \Comment{Eqs.~\eqref{eq:u_def}--\eqref{eq:v_def}, \eqref{eq:uhat_vhat}}
  \State \(\widehat C_\sigma^{\mathrm{num}}(N_m)\gets\widehat u_m+\sigma\widehat v_m\); \(\widehat C_\sigma(N_m)\gets\widehat C_\sigma^{\mathrm{num}}(N_m)/[1+\sigma\widehat r_{ZI}(N_m)]\); demodulate to \(\widetilde{C}_\sigma(N_m)\) \Comment{Eqs.~\eqref{eq:C_branch_num}, \eqref{eq:C_branch}, \eqref{eq:C_demod}}
\EndFor
\State \textbf{Estimate} \(\widehat\vartheta_c\) from \(\{\widetilde{C}_\pm(N_m)\}\) via the matrix-pencil branch-pole estimator [Algorithm~\ref{alg:pencil}]; the analytic branch-linear slope estimate \(\widehat\vartheta_c=(s_+-s_-)/2\) [Eqs.~\eqref{eq:unwrap_phase}--\eqref{eq:vartheta_hat_branchlin}] is its transparent linear-phase limit\ 
\State \Return \(\widehat\vartheta_c\) \Comment{optionally map to \(\delta_{\mathrm{CP}}\) via Eq.~\eqref{eq:cp_vs_zz_angle}}
\end{algorithmic}
\end{figure}

\subsection{Scope and assumptions}
\label{par:scope-limits}
Branch-resolved Pauli-block spectroscopy as developed above targets a single coherent residual \(ZZ\) rotation per calibrated cycle in the normal form of Eq.~\eqref{eq:normal_form}.
It assumes that contrast loss is well described by a single effective damping rate \(\Gamma\), that leakage and higher-order coherent faults remain subdominant over the sampled sequence lengths, and that control-qubit \(ZI\) drift is negligible or removable by the fixed analysis frame.
Two failure modes follow directly from these assumptions.
First, if leakage to non-computational levels induces multi-frequency beating in the branch sequences, the single-pole (single-slope) model underlying Eqs.~\eqref{eq:C_ideal}--\eqref{eq:vartheta_hat_branchlin} no longer holds, and a multi-tone or leakage-aware readout is required.
Second, if control-qubit \(ZI\) drift is not negligible, the unnormalized numerators \(C_\sigma^{\mathrm{num}}\) no longer coincide with the conditional branch coherences, and readout must retain the normalization denominator \(1+\sigma r_{ZI}\) of Eq.~\eqref{eq:C_branch} rather than using \(C_\sigma^{\mathrm{num}}\) directly.\ 
The closed block \(\{IX,IY,ZX,ZY\}\) is required whenever \(\beta_c\neq 0\); a two-moment sector readout is adequate only in the special case \(\beta_c\simeq 0\).
An echoed-cycle template [Eq.~\eqref{eq:echo_cycle}] may substitute for \(U_{\mathrm{cyc}}\) when removable \(\beta_c\) should be suppressed before readout (Sec.~\ref{subsec:numerical-validation}).
Frame bookkeeping---virtual-\(Z\) updates, echo sign flips, and carrier demodulation---must be applied before averaging; uncorrected toggling erases branch coherence even when the underlying \(ZZ\) term survives.
The estimator returns a signed per-cycle angle \(\widehat\vartheta_c\) suitable for pulse update, not a full process matrix, channel spectrum, or guaranteed convergence of multi-round feedback.
Mapping to \(\delta_{\mathrm{CP}}\) follows Eq.~\eqref{eq:cp_vs_zz_angle} once the hardware sign convention is fixed.
Sec.~\ref{sec:results} tests these claims under injected noise and on one production qubit--coupler pair; extension to automated actuator selection, leakage-aware readout, and systematic multi-pair deployment remains outside the present scope.

\section{Results and Discussion}
\label{sec:results}

Section~\ref{sec:theory} developed branch-resolved Pauli-block spectroscopy and stated the operational claims to be tested; we now ask whether signed per-cycle estimates of the residual \(ZZ\) angle \(\widehat\vartheta_c\) follow within that scope.
Numerical circuit-protocol simulations with injected phase, detuning, damping, and SPAM test the branch readout chain on the bare cycle of Fig.~\ref{fig:protocol_circuit}(a); experiments on a superconducting cloud processor implement the echoed supercycle of panel~(b) on native \hwcode{CZ} pulses in one qubit--coupler session.
Figures~\ref{fig:scalar_vs_branch}--\ref{fig:complementarity} proceed from scalar-versus-branch necessity through injected linearity, operational robustness, an effective echoed-cycle check, and complementarity with fidelity; Fig.~\ref{fig:guodun_hardware} closes the loop with pulse-level injection and feedback on the literal echoed template.

\subsection{Numerical validation}
\label{subsec:numerical-validation}

\begin{figure}[t]
    \centering
    \includegraphics[width=\columnwidth]{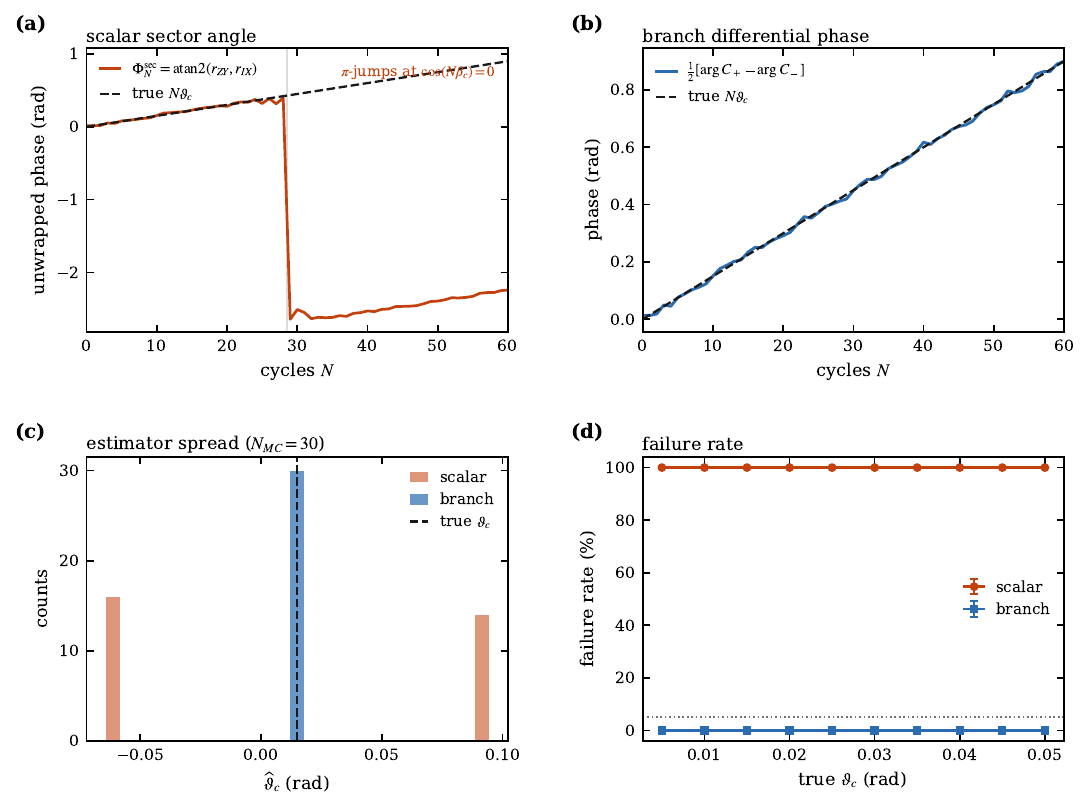}
    \caption{Scalar- versus branch-resolved readout at \(\beta_c\neq 0\) (full circuit-protocol simulation).
    (a)~Scalar angle \(\Phi_N^{\mathrm{sec}}=\operatorname{atan2}(r_{ZY},r_{IX})\) develops a \(\pi\)-jump near \(\cos(N\beta_c)=0\).
    (b)~Branch differential phase \(\tfrac{1}{2}[\arg C_+-\arg C_-]\) tracks injected \(\vartheta_c\) linearly in~\(N\).
    (c)~Estimator spread over \(N_{\mathrm{MC}}=30\) independent trials at fixed \(\vartheta_c\).
    (d)~Failure rate versus injected \(\vartheta_c\); branch readout remains reliable where scalar fails.
    Each trial prepares \(\ket{+}\ket{+}\) [Eq.~\eqref{eq:probe_state}], evolves \(N=0,\ldots,N_{\max}\) cycles of Eq.~\eqref{eq:normal_form}, and measures the two target-basis settings of Sec.~\ref{subsec:protocol}, decoding the four Pauli moments, by multinomial shot sampling (\(S=2\times10^4\) shots per basis setting) from joint computational outcomes, followed by a fixed additive SPAM offset \(b_{\mathrm{SPAM}}=10^{-4}\) on each \(\widehat r_P(N)\) (\(b_{\mathrm{SPAM,mult}}=0\)).\ 
    Contrast damping \(D_N=Ae^{-\Gamma N}\) with \(A=0.95\) and \(\Gamma=10^{-3}\)~cycle\(^{-1}\) is applied to coherences before sampling.
    Panels (a)--(c) use \(\beta_c=0.055\), \(\vartheta_c=0.015\)~rad/cycle, \(\omega_0=0\), and \(N_{\max}=60\); panel~(d) sweeps \(\vartheta_c\in[0.005,0.050]\)~rad (\(10\) values) with \(N_{\mathrm{MC,sweep}}=40\) trials per point.
    From the \emph{same} simulated \(\{\widehat r_P(N)\}\) in each trial, the scalar estimate is the weighted linear slope of \(\mathrm{unwrap}\,\Arg(r_{IX}+\ii r_{ZY})\) versus \(N\), whereas the branch estimate is \(\widehat\vartheta_c=(s_+-s_-)/2\) from unwrapped \(\Arg C_\pm(N)\) [Eqs.~\eqref{eq:unwrap_phase}--\eqref{eq:vartheta_hat_branchlin}]; both panel traces apply phase unwrapping before slope fitting.
    A trial is counted as failed when \(|\widehat\vartheta_c-\vartheta_c|>\max(0.25|\vartheta_c|,10^{-3})\)~rad.  
    }
    \label{fig:scalar_vs_branch}
\end{figure}

We implement the protocol of Sec.~\ref{subsec:protocol} as a numerical circuit-protocol simulation rather than by evaluating closed-form Pauli trajectories alone.
Each trial prepares the product probe \(\ket{+}\ket{+}\) [Eq.~\eqref{eq:probe_state}], applies \(N\) repetitions of the bare residual cycle [Fig.~\ref{fig:protocol_circuit}(a)]
\(U_{\mathrm{cyc}}=\exp[-\mathrm{i}(\beta_c IZ+\vartheta_c ZZ)/2]\),
measures the two target-basis settings \(R_X,R_Y\) and decodes the four Pauli moments \(\{IX,IY,ZX,ZY\}\) from joint computational readout [Eq.~\eqref{eq:pauli_from_counts}], and forms \(\widehat r_P(N)\) from sampled bit strings.\ 
Contrast damping \(D_N=Ae^{-\Gamma N}\) and additive SPAM offsets on the Pauli moments are included with the parameters listed in the caption of Fig.~\ref{fig:scalar_vs_branch}.
Each histogram or failure rate aggregates independent trials at fixed noise parameters.

Section~\ref{sec:theory-pauli} predicts that a two-moment readout in the \(\{IX,ZY\}\) sector fails once \(\beta_c\neq 0\) because the neglected \(IY\) and \(ZX\) components enter through \(\cos B_N\) in Eqs.~\eqref{eq:rIX_full}--\eqref{eq:rZY_full} [Eq.~\eqref{eq:BN_ThetaN}].
We therefore compare that scalar sector readout with branch-resolved estimation under injected \(\beta_c\neq 0\) using the same simulated bit-string data.
The scalar angle \(\Phi_N^{\mathrm{sec}}=\operatorname{atan2}[r_{ZY}(N),r_{IX}(N)]\) tracks the amplified rotation when \(\beta_c=0\), but for \(\beta_c\neq 0\) the \(\cos B_N\) envelope forces \(\pi\)-discontinuities in \(\Phi_N^{\mathrm{sec}}\) whenever the effective quadrature changes sign [Fig.~\ref{fig:scalar_vs_branch}(a)], as anticipated above.
With the full block \(\{IX,IY,ZX,ZY\}\) and branch coherences \(C_\pm\) [Eqs.~\eqref{eq:C_branch_num}, \eqref{eq:C_branch}], the differential branch phase instead evolves smoothly and remains linear in the accumulated residual angle [panel~(b)].
Over thirty independent trials at fixed injected \(\vartheta_c\), the scalar estimate is broadly dispersed and biased away from the injected value, whereas branch-resolved \(\widehat\vartheta_c=(s_+-s_-)/2\) from Eq.~\eqref{eq:vartheta_hat_branchlin} concentrates on the injected value [panel~(c)].
In the injected-angle sweep of panel~(d), scalar readout fails in every trial across the range shown, whereas branch readout maintains zero failures under the stated acceptance criterion.
These simulations confirm the need for the full Pauli block when \(\beta_c\neq 0\) and support the branch-resolved construction of Sec.~\ref{sec:theory-pauli}.

\begin{figure*}[t]
    \centering
    \includegraphics[width=0.98\textwidth]{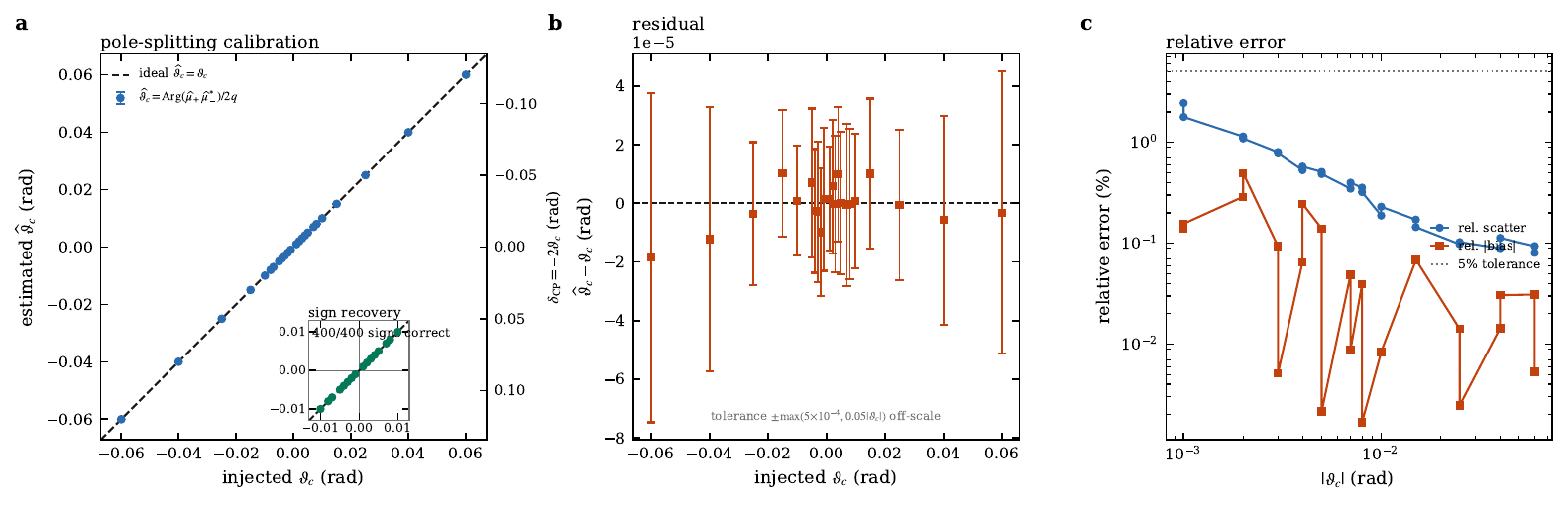}
    \caption{Injected-phase calibration of branch-resolved \(\widehat\vartheta_c\) (matrix-pencil readout).
    (a)~\(\widehat\vartheta_c\) versus injected \(\vartheta_c\); dashed line, unit slope; inset, sign recovery near zero.
    (b)--(c)~Estimation residual and relative error versus \(|\vartheta_c|\) (5\% tolerance in~(c)).
    \(24\) injection angles in \([-0.06,0.06]\)~rad/cycle, with the smallest tested magnitude \(|\vartheta_c|=10^{-3}\)~rad/cycle; \(N_{\mathrm{MC}}=25\) trials each; \(\beta_c=0.03\), \(S=1.5\times10^4\) shots per Pauli setting; probe \(\ket{+}\ket{+}\) [Sec.~\ref{subsec:protocol}].
    Error bars in (a) are the across-trial standard deviation and set the empirical resolution floor near zero; at the smallest tested magnitude the sign of \(\vartheta_c\) is recovered in all \(N_{\mathrm{MC}}\) trials.  
    }
    \label{fig:injected_phase_linearity}
\end{figure*}

\begin{figure}[t]
    \centering
    \includegraphics[width=\columnwidth]{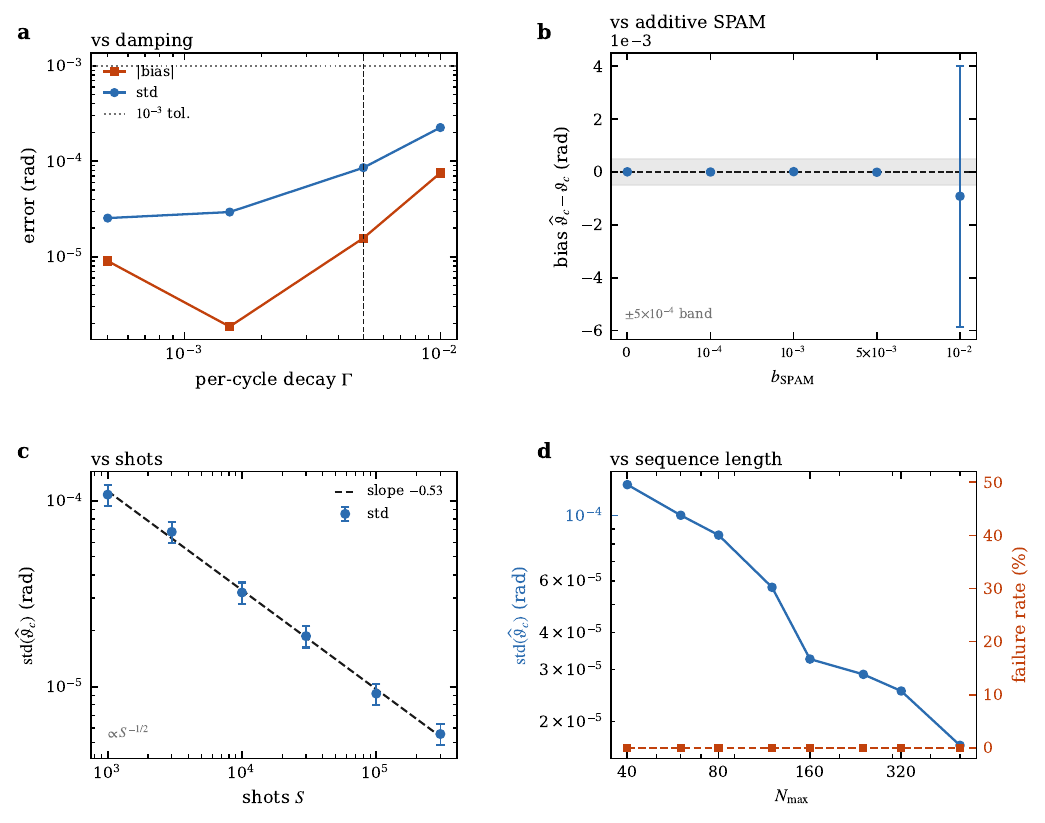}
    \caption{Robustness of branch-resolved \(\widehat\vartheta_c\) estimation (matrix-pencil readout).
    (a)~Bias and std versus per-cycle decay \(\Gamma\).
    (b)~Bias versus SPAM offset \(b_{\mathrm{SPAM}}\).
    (c)~Std versus shots \(S\) (\(\propto S^{-1/2}\)).
    (d)~Std and failure rate versus \(N_{\max}\).
    Fixed \(\beta_c=0.035\), \(\vartheta_c=0.02\)~rad/cycle; baseline \(S=1.5\times10^4\), \(N_{\max}=320\), \(N_{\mathrm{MC}}=30\); probe \(\ket{+}\ket{+}\) [Sec.~\ref{subsec:protocol}].}
    \label{fig:noise_robustness_branch}
\end{figure}

With the full block validated at \(\beta_c\neq 0\), we next ask whether the branch-resolved estimator of Eq.~\eqref{eq:vartheta_hat_branchlin} returns the injected per-cycle angle when \(\vartheta_c\) is swept through the same circuit protocol of Sec.~\ref{subsec:protocol}.
Equations~\eqref{eq:C_demod}--\eqref{eq:vartheta_hat_branchlin} predict an unbiased, approximately unit-slope response in \(\widehat\vartheta_c\) over a controlled injection range, including signed recovery near \(\vartheta_c=0\).
Consistent with Sec.~\ref{subsec:protocol}, readout in Figs.~\ref{fig:injected_phase_linearity} and \ref{fig:noise_robustness_branch} uses the matrix-pencil pole extractor of Algorithm~\ref{alg:pencil}; the analytic branch-linear readout [Eq.~\eqref{eq:vartheta_hat_branchlin}] yields equivalent estimates within statistical scatter in the linear-phase limit (Fig.~\ref{fig:app_readout_comparison}).\ 

Figure~\ref{fig:injected_phase_linearity}(a) plots \(\widehat\vartheta_c\) versus injected \(\vartheta_c\) over \(\pm 0.06\)~rad/cycle, with a secondary axis for the corresponding controlled-phase residual \(\delta_{\mathrm{CP}}=-2\vartheta_c\) under Eq.~\eqref{eq:cp_vs_zz_angle}; the inset zooms on \(|\vartheta_c|\lesssim 0.012\)~rad/cycle and confirms sign recovery at every trial, down to the smallest tested magnitude \(|\vartheta_c|=10^{-3}\)~rad/cycle where all \(N_{\mathrm{MC}}\) trials recover the correct sign.
The empirical resolution floor near zero is set by the across-trial estimator scatter (error bars), which is the relevant confidence interval at fixed shot budget rather than a hard threshold.\ 
The estimation residuals [panel~(b)] and relative errors [panel~(c)] remain centered on zero with magnitude consistent with the simulation scatter.

We then stress the same circuit-protocol branch-resolved pipeline under the operational degradations allowed in the scope of Sec.~\ref{par:scope-limits}: contrast damping \(\Gamma\) in \(D_N=Ae^{-\Gamma N}\), additive SPAM offsets on the Pauli moments, finite shot count \(S\), and maximum sequence length \(N_{\max}\), while holding the injected \(\vartheta_c\) fixed.

Figure~\ref{fig:noise_robustness_branch} sweeps these four parameters in turn on bit-string data generated as in Sec.~\ref{subsec:protocol}.
Per-cycle damping attenuates branch-pole moduli as expected from \(D_N\), yet leaves \(|\widehat\vartheta_c-\vartheta_c|\) below the \(10^{-3}\)~rad tolerance over the tested \(\Gamma\) range [panel~(a)].
Additive SPAM offsets shift individual Pauli moments but produce no systematic bias in \(\widehat\vartheta_c\) within the shaded \(\pm 5\times 10^{-4}\)~rad band [panel~(b)].
This SPAM robustness is a property of the working matrix-pencil estimator, which isolates the branch signal poles from the DC and nuisance components carrying the offset, rather than an automatic immunity of a bare branch-linear phase fit; the estimator-comparison stress tests of Appendix~\ref{app:readout-comparison} (Fig.~\ref{fig:app_readout_comparison}) quantify this advantage under branch-asymmetric SPAM.\ 
The estimator standard deviation follows the expected \(S^{-1/2}\) scaling with shot count [panel~(c)] and decreases with \(N_{\max}\) without readout failures [panel~(d)].
Together with Fig.~\ref{fig:injected_phase_linearity}, these sweeps indicate that signed, unbiased \(\widehat\vartheta_c\) readout persists once the full Pauli block handles \(\beta_c\).

\begin{figure}[t]
    \centering
    \includegraphics[width=\columnwidth]{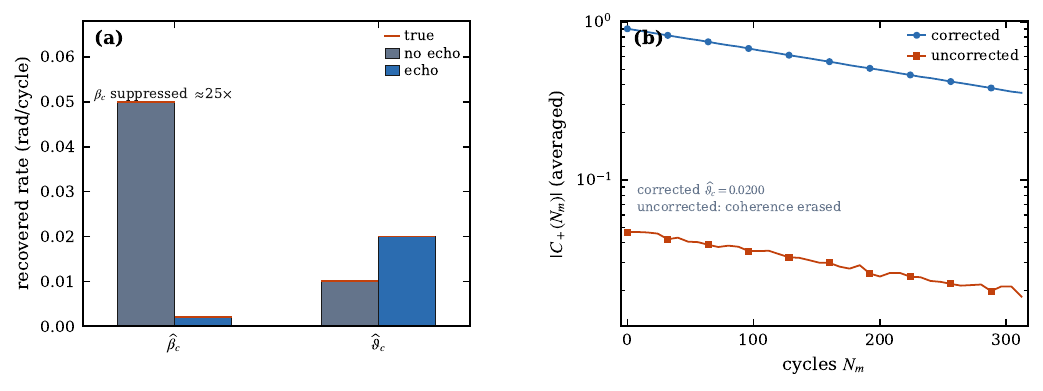}
    \caption{Echoed-cycle variant and fixed-frame correction (matrix-pencil readout).
    This is an effective echoed-cycle simulation: echo is modeled by imposed effective per-cycle \((\beta_c,\vartheta_c)\) pairs reflecting the toggling-frame action of Eq.~\eqref{eq:echo_cycle}, not by literal \((XX)\) pulses around a physical \(U_{\mathrm{cyc}}\).\ 
    All rates are per repeated cycle; for the echo condition this is one echoed cycle [Eq.~\eqref{eq:echo_cycle}, two entangling blocks].\ 
    (a)~\(\widehat\beta_c\) and \(\widehat\vartheta_c\) with and without echo [imposed \((\beta_c,\vartheta_c)=(0.05,0.01)\) vs.\ \((0.002,0.02)\)~rad/cycle].
    (b)~Mean \(|C_+(N_m)|\) after random frame averaging, with and without offline sign correction.
    \(S=1.5\times10^4\) shots, \(N_{\max}=320\), \(N_{\mathrm{MC}}=30\) [(a)], \(N_{\mathrm{frame}}=150\) [(b)]; probe \(\ket{+}\ket{+}\) [Sec.~\ref{subsec:protocol}].}
    \label{fig:optional_echo}
\end{figure}

An optional echoed template [Fig.~\ref{fig:protocol_circuit}(b), Eq.~\eqref{eq:echo_cycle}] replaces each bare cycle by
\begin{equation}
U_{\mathrm{cyc}}^{\mathrm{echo}}
=
(XX)\,U_{\mathrm{cyc}}\,(XX)\,U_{\mathrm{cyc}},
\label{eq:echo_cycle}
\end{equation}
where \((XX)\) denotes a simultaneous \(X\) rotation on both qubits, i.e., a dynamical-decoupling refocus pulse~\cite{Viola1999DynamicalDecoupling} analogous to echoed two-qubit calibration~\cite{Sundaresan2020HEAT}.
In the toggling frame of the refocus pulse,
\begin{align}
(XX)(IZ)(XX)&=-IZ,\\
(XX)(ZI)(XX)&=-ZI,\\
(XX)(ZZ)(XX)&=ZZ,
\label{eq:echo_toggling}
\end{align}
so leading local \(IZ\) and \(ZI\) contributions reverse across each \((XX)\) sandwich and cancel over Eq.~\eqref{eq:echo_cycle}, whereas the nonlocal \(ZZ\) term is preserved and accumulates across the two entangling blocks.\ 
Each echoed cycle in Eq.~\eqref{eq:echo_cycle} contains two entangling blocks; in the fixed analysis frame the refocusing suppresses removable per-cycle \(IZ\) accumulation while the \(ZZ\) contributions of the two blocks add.
Consequently, when the echoed template is used, the cycle index \(N\) counts echoed cycles rather than individual entangling gates, and the estimator returns the residual \(ZZ\) angle \(\vartheta_c\) accumulated \emph{per echoed cycle}---the same unit used throughout for injection calibration, feedback update, and the mapping \(\delta_{\mathrm{CP}}=-2\vartheta_c\) of Eq.~\eqref{eq:cp_vs_zz_angle}; a per-single-gate value, if desired, would follow by dividing by the number of entangling blocks per cycle.
The readout and estimator of Sec.~\ref{subsec:protocol} are otherwise unchanged once the corresponding frame signs are corrected offline.
Figure~\ref{fig:optional_echo} evaluates this template as an \emph{effective} echoed-cycle simulation: rather than compiling literal \((XX)\) refocusing pulses around a physical \(U_{\mathrm{cyc}}\), native and echoed cycles are represented in the same bit-string circuit-protocol pipeline as Figs.~\ref{fig:injected_phase_linearity} and \ref{fig:noise_robustness_branch} by distinct effective per-cycle \((\beta_c,\vartheta_c)\) pairs, chosen to reflect the toggling-frame action of Eq.~\eqref{eq:echo_cycle} (leading \(IZ\) refocused, \(ZZ\) preserved).\ 
Panel~(a) therefore does not derive echo suppression ab initio; it imposes the reduced effective \(\beta_c\) and verifies that the branch readout resolves the resulting \(\widehat\beta_c\) suppression (here by roughly twenty-five-fold) while \(\widehat\vartheta_c\) still tracks the injected value.
Panel~(b) shows that branch coherence is erased if random frame signs are not corrected before averaging, but survives over hundreds of cycles once the analysis frame is fixed.

\begin{figure}[t]
    \centering
    \includegraphics[width=\columnwidth]{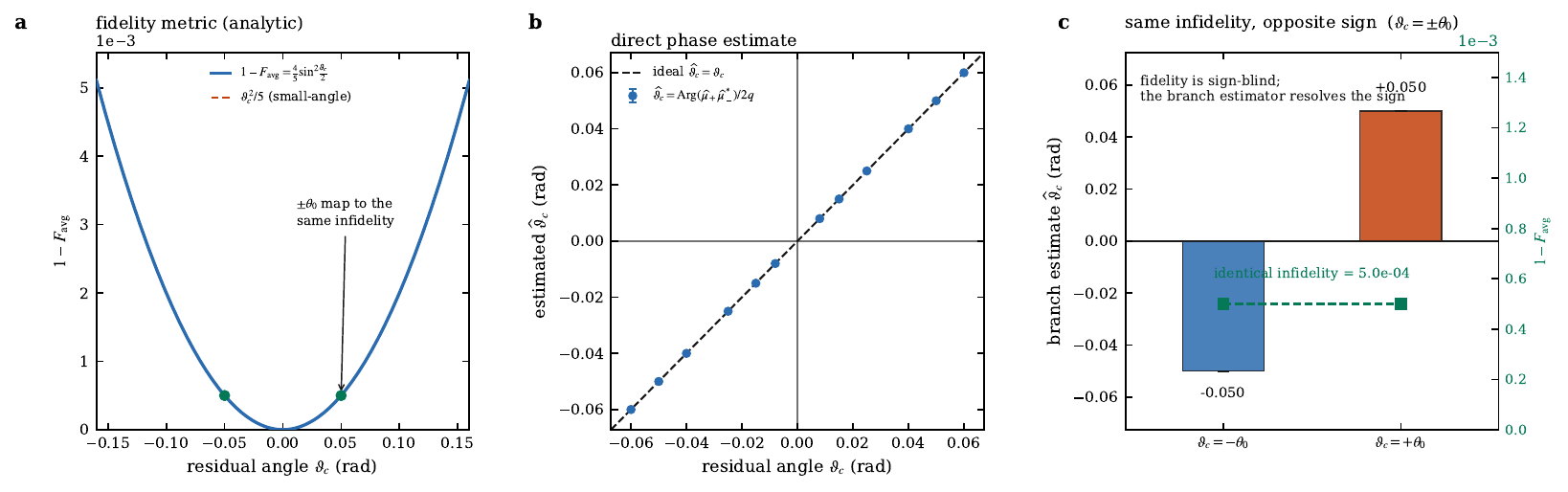}
    \caption{Complementarity of average gate infidelity and signed \(\widehat\vartheta_c\) estimation (matrix-pencil readout).
    (a)~Analytic \(1-F_{\mathrm{avg}}\) for \(\exp(-\ii\vartheta_c ZZ/2)\); even in sign.
    (b)~\(\widehat\vartheta_c\) versus injected \(\vartheta_c\).
    (c)~At \(\vartheta_c=\pm\vartheta_0=0.05\)~rad/cycle, opposite estimates with identical infidelity.\ 
    \(12\) injection angles, \(N_{\mathrm{MC}}=25\) trials each; probe \(\ket{+}\ket{+}\) [Sec.~\ref{subsec:protocol}]; panel~(a) is an analytic reference.}
    \label{fig:complementarity}
\end{figure}

Finally, average gate infidelity of a residual unitary \(\exp(-\ii\vartheta_c ZZ/2)\) scales as \(\mathcal{O}(\vartheta_c^2)\) for small angles and is unchanged under \(\vartheta_c\to-\vartheta_c\).
Figure~\ref{fig:complementarity} evaluates this contrast in the same bit-string circuit-protocol simulation as Figs.~\ref{fig:injected_phase_linearity}--\ref{fig:optional_echo}.
Panel~(a) plots the analytic average-gate infidelity \(1-F_{\mathrm{avg}}=\frac{4}{5}\sin^2(\vartheta_c/2)\), which is even in sign and therefore blind to the residual rotation direction.
Panel~(b) shows that the branch-pole readout returns a signed, approximately linear estimate \(\widehat\vartheta_c\) across the injection grid.
Panel~(c) highlights the complementarity at matched magnitude \(\vartheta_c=\pm\vartheta_0\): the fidelity reference is identical, whereas \(\widehat\vartheta_c\) reverses sign.\ 
The direct readout therefore supplies calibration information that fidelity-oriented benchmarks suppress at leading order; the fidelity curve in panel~(a) is an analytic reference rather than an independent randomized-benchmarking experiment.

\subsection{Hardware validation on a superconducting processor}
\label{subsec:hardware-results}

The numerical study establishes signed branch readout under injected noise; the remaining question is whether the same readout and one-step correction transfer to native two-qubit pulses, where each cycle is compiled from calibrated \hwcode{CZ} waveforms and the residual phase must be actuated through hardware knobs rather than an ideal generator. We therefore deploy the full closed loop on a production superconducting cloud processor.\ 

We use the QuantumCTek cloud processor \hwcode{gd\_qc1}, coupler G55 (control Q36, target Q30), selected from a 12-pair echoed-cycle survey by branch contrast (selection \minC\ $\approx 0.99$ on G55; platform, pulse, and readout details in Appendix~\ref{app:hardware-experiment}).\ 
With both qubits initialized in \(\ket{+}\) [Eq.~\eqref{eq:probe_state}], we run the echoed supercycle of Fig.~\ref{fig:protocol_circuit}(b) [Eq.~\eqref{eq:echo-cycle-hw}] on GetPulse-calibrated \hwcode{CZ} waveforms, which refocuses removable \(\beta_c\) before branch readout, and close the loop in four native steps: (i)~survey pulse actuators for a knob that moves \(\vartheta_c\) without collapsing the \(\{IX,IY,ZX,ZY\}\) block; (ii)~calibrate its gain \(G_{\mathrm{inj}}\) against the differential branch slope; (iii)~verify injected-phase linearity in \(\vartheta_{\mathrm{inj}}\) as predicted by Eqs.~\eqref{eq:C_demod}--\eqref{eq:vartheta_hat_branchlin}; and (iv)~apply one signed feedback step.
Logical \hwcode{CZ}/\hwcode{RZ} injection did not yield a unit-slope map on this backend, so the reported data use pulse-native detune control through \paramqzerodetune.

\begin{figure}[t]
    \centering
    \includegraphics[width=\columnwidth]{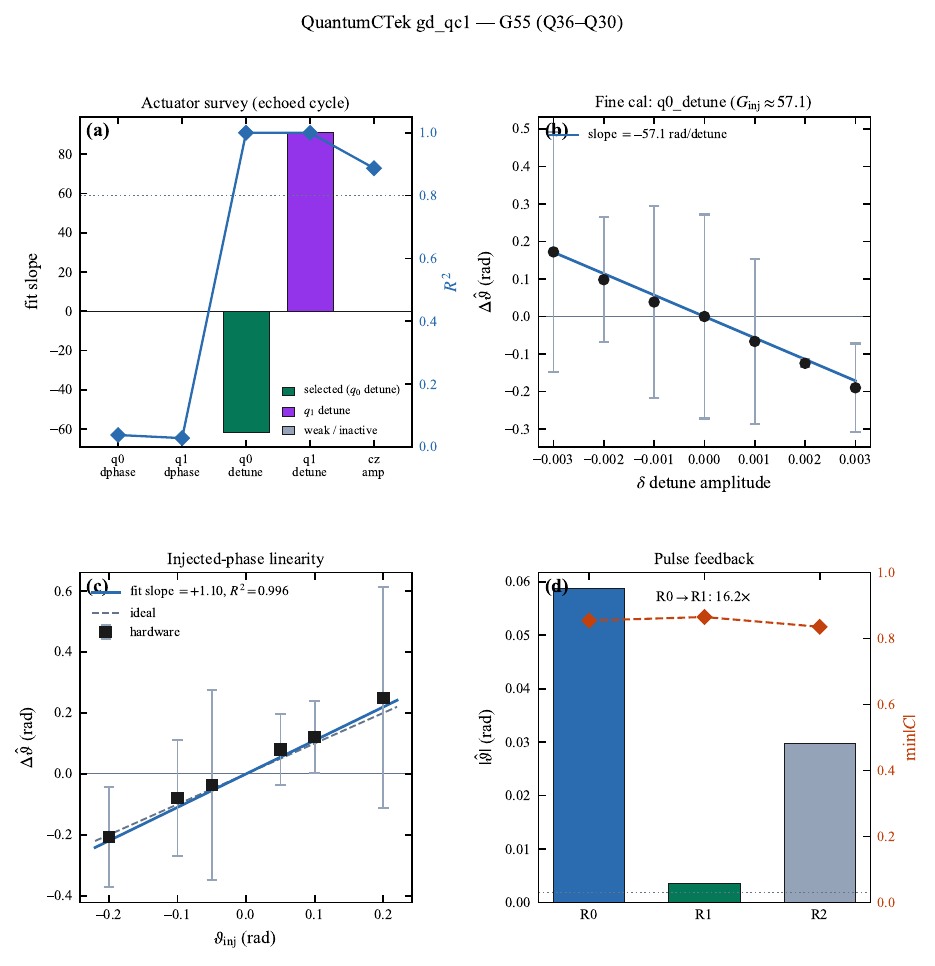}
    \caption{
    Hardware validation on \hwcode{gd\_qc1}, coupler G55 (Q36--Q30).
    All \(\widehat\vartheta_c\) values are per echoed cycle [Eq.~\eqref{eq:echo-cycle-hw}, two native \hwcode{CZ} blocks].\ 
    (a)~Actuator survey on echoed cycles: fit slope (bars) and \(R^2\) (markers) for five GetPulse knobs; \paramqzerodetune\ preserves branch contrast (\minC\ $\approx 0.86$).
    (b)~Fine gain calibration of \(G_{\mathrm{inj}}\): $G_{\mathrm{inj}} \approx 57.1$~rad/(detune unit), $R^2 \approx 0.998$.\ 
    (c)~Injected-phase linearity versus $\vartheta_{\mathrm{inj}} \in \{0, \pm 0.05, \pm 0.1, \pm 0.2\}$~rad: slope $1.10$, $R^2 = 0.996$.\ 
    (d)~One pulse-feedback step yields a point-estimate ratio $|\widehat\vartheta_c|_{\mathrm{R0}}/|\widehat\vartheta_c|_{\mathrm{R1}}\approx 16$ (full-precision value $16.2$) while \minC\ remains $\gtrsim 0.83$ (bootstrap std in Table~\ref{tab:closed-loop}).
    }
    \label{fig:guodun_hardware}
\end{figure}

Figure~\ref{fig:guodun_hardware} reports the outcome.
The selected actuator preserves branch contrast (\minC\ $\approx 0.86$) while giving a clean linear gain \(G_{\mathrm{inj}}\approx 57.1\)~rad per detune unit ($R^2 \approx 0.998$) [panels~(a),(b)]; the injected-phase response is near-linear, with slope $1.10$ and $R^2 = 0.996$ across \(\vartheta_{\mathrm{inj}}\in\{0,\pm 0.05,\pm 0.1,\pm 0.2\}\)~rad [panel~(c)], matching the unit-slope prediction of Eqs.~\eqref{eq:C_demod}--\eqref{eq:vartheta_hat_branchlin}.
Applying one pulse-feedback step from the signed estimate suppresses the native residual by a point-estimate ratio \(|\widehat\vartheta_c|_{\mathrm{R0}}/|\widehat\vartheta_c|_{\mathrm{R1}}\approx 16\) (full-precision value \(16.2\)) while branch contrast remains \minC\ $\gtrsim 0.83$ [panel~(d), Table~\ref{tab:closed-loop}].\ 

Together these panels show that the full chain---echoed cycles, actuator selection, injection calibration, signed branch readout, and one-step correction---runs on a production cloud processor for one qubit--coupler pair in a single session.
Because the six-depth hardware bootstrap distribution is broad and heavy-tailed, the feedback result is a sign-consistent point-estimate suppression rather than a statistically precise residual measurement.\ 
With the numerical validation above, this establishes branch-resolved metrology as a practical signed update variable for native conditional-phase calibration rather than a standalone fit report; extension to multiple pairs and convergent multi-round feedback with higher statistics is left for future work.

\section{Conclusion}
\label{sec:conclusion}

We have presented branch-resolved Pauli-block spectroscopy as repeated-cycle metrology of the signed per-cycle residual \(ZZ\) angle that governs small coherent conditional-phase drift in calibrated two-qubit cycles.
Reconstruction of the closed Pauli block \(\{IX,IY,ZX,ZY\}\) and differential readout of branch coherences \(C_\pm\) isolate nonlocal \(\vartheta_c\) from target-qubit detuning \(\beta_c\); echoed cycles and fixed-frame bookkeeping suppress removable local accumulation before estimation.

Circuit-protocol simulations yield unbiased, sign-resolving branch estimates under injected detuning, contrast damping, SPAM, and finite sampling---regimes where scalar-sector readout fails and leading-order infidelity is blind to rotation direction---while on native CZ pulses of a cloud superconducting processor, pulse-level injection and one-step point-estimate feedback preserve branch contrast on one production qubit--coupler pair.\ 

The method therefore supplies a signed per-cycle control variable directly from a fixed four-moment Pauli block, closing the loop between repeated-cycle spectroscopy and native pulse correction of conditional-phase error without full channel tomography or sign-degenerate fidelity benchmarking.

\section*{Acknowledgements}

\noindent\textbf{Funding:}
This work is supported by the National Natural Science Foundation of China (grant no.~92365111), Shanghai Municipal Science and Technology (grant no.~25LZ2600200), the Beijing Natural Science Foundation (grant no.~Z220002), the Quantum Science and Technology-National Science and Technology Major Project, QNMP (Grant No.~2021ZD0302400), and the National Key Research and Development Program of China (grant no.~2025YFE0217200).

\noindent\textbf{Author contributions:}
Xudan Chai conceived the project, performed the simulations and experiments, and wrote the manuscript; Yanwu Gu and Dong E. Liu contributed to discussions; Kerui Li and Huiqi Xue contributed to the experiments.
We thank Dr.~Bo Gao at the Beijing Institute of Technology for careful revision of the manuscript.

\noindent\textbf{Competing interests:}
The authors declare that they have no competing interests.

\noindent\textbf{Data and code availability:}
The data and code supporting this study will be openly released in a public repository upon publication and are available from the corresponding author on reasonable request.

\FloatBarrier
\bibliography{refs}
\newpage

\appendix
\label{appendix}

\section{Branch-resolved readout and estimator comparison}
\label{app:readout-comparison}

\begin{figure}[b]
    \centering
    \includegraphics[width=\columnwidth]{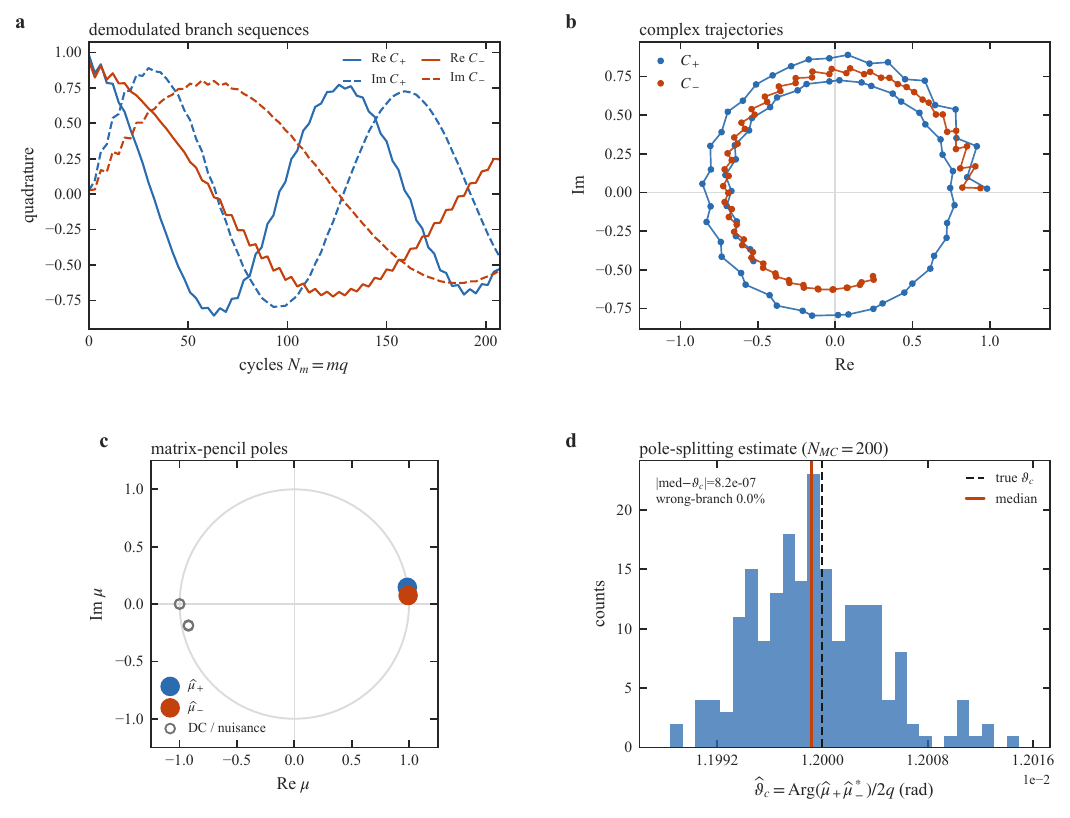}
    \caption{Matrix-pencil readout on demodulated branch coherences.
    (a)~Real and imaginary parts of \(\widetilde{C}_\pm(N_m)\) versus cycle index \(N_m=mq\).
    (b)~Complex trajectories of \(\widetilde{C}_\pm\) in the measurement frame.
    (c)~Selected non-DC sampling-step poles \(\widehat\lambda_\pm\) in the complex plane (unit circle shown); open circles, DC and nuisance components.
    (d)~Pole-splitting estimate \(\widehat\vartheta_c=\Arg(\widehat\lambda_+\widehat\lambda_-^{*})/(2q)\) over $N_{\mathrm{MC}}=200$ independent trials at fixed injected \(\vartheta_c\).}\ 
    \label{fig:pencil_branch_poles}
\end{figure}

Section~\ref{sec:theory} introduces the branch-linear readout of Eq.~\eqref{eq:vartheta_hat_branchlin} as the transparent analytic estimator that makes the phase structure explicit, and adopts the matrix-pencil pole readout described here as the working estimator under realistic shot noise, damping, and SPAM.\ 
Accordingly, the numerical validations in Figs.~\ref{fig:injected_phase_linearity}, \ref{fig:noise_robustness_branch}, \ref{fig:optional_echo}, and \ref{fig:complementarity} implement the matrix-pencil pole readout on demodulated branch sequences \(\widetilde{C}_\pm(N)\), which reduces to Eq.~\eqref{eq:vartheta_hat_branchlin} within statistical scatter in the linear-phase limit.
Figure~\ref{fig:pencil_branch_poles} summarizes that pipeline: after carrier demodulation, a matrix-pencil analysis is applied separately to each branch; the selected non-DC sampling-step poles \(\widehat\lambda_\pm\) encode the branch phases, and the per-cycle angle follows from the differential pole phase \(\Arg(\widehat\lambda_+\widehat\lambda_-^{*})/(2q)\).\ 
Algorithm~\ref{alg:pencil} specifies the Hankel dimensions, pole selection, validity gate, and bootstrap resampling used in all reported numerical and hardware analyses; it takes the measured depths \(N_m\) as input.
The depth grid itself is a data-acquisition choice, not part of the estimator: in the adaptive-grid simulations we set \(N_{\max}=\mathrm{clip}(\lceil 8/|\vartheta_c^{\mathrm{guess}}|\rceil,N_{\mathrm{lo}},N_{\mathrm{hi}})\) from an initial angle-scale estimate and sample \(N_m=mq\) with \(q=\max(1,\mathrm{round}[N_{\max}/(N_s-1)])\), whereas the hardware runs use a fixed short grid \(N=1,\ldots,6\) [Appendix~\ref{app:hardware-experiment}].\ 

\begin{figure}[t]
\refstepcounter{algorithmcounter}\label{alg:pencil}
\noindent\textbf{Algorithm~\thealgorithmcounter.} Matrix-pencil branch-pole estimator
\vspace{0.35em}
\begin{algorithmic}[1]
\Require Demodulated branch sequences \(\{\widetilde{C}_+(N_m),\widetilde{C}_-(N_m)\}\) at cycle depths \(N_m=N_0+mq\), \(m=0,\ldots,N_s-1\); subsample period \(q\); pencil order \(K=3\); Hankel row count \(L=\lfloor N_s/2\rfloor\); DC tolerance \(\delta_{\mathrm{DC}}=5\times10^{-3}\); contrast threshold \(c_{\min}\) (0.03 in hardware); pole-modulus threshold \(\lambda_{\min}=0.30\); bootstrap count \(B\) (200 in hardware)
\Ensure Per-cycle estimate \(\widehat\vartheta_c\); optional bootstrap scatter \(\widehat\sigma_{\mathrm{boot}}\)
\For{each branch \(\sigma\in\{+,-\}\)}
  \State Form \(y_{\sigma,m}\gets\widetilde{C}_\sigma(N_m)\), \(m=0,\ldots,N_s-1\) \Comment{constant offset \(N_0\) is absorbed into the amplitudes and does not affect the pole phase}\ 
  \State Build Hankel matrices \(Y_0,Y_1\in\mathbb{C}^{L\times(N_s-L)}\) from \(\{y_{\sigma,m}\}\), \(L=\lfloor N_s/2\rfloor\)
  \State SVD \(Y_0=USV^\dagger\); truncate to \(K_{\mathrm{eff}}=\min\{K,L,N_s-L,\#\{s_i>s_1\times10^{-9}\}\}\)
  \State \(A_p\gets (U_K^\dagger Y_1 V_K)\,\mathrm{diag}(S_K)^{-1}\); sampling-step poles \(\{\lambda_j\}\gets\mathrm{eig}(A_p)\)\ 
  \State Fit \(y_{\sigma,m}\approx\sum_j a_j\lambda_j^{m}+c_{\mathrm{DC}}\) by least squares, projecting \(\lambda_j\to\lambda_j/|\lambda_j|\) when \(|\lambda_j|>1\)
  \State Reject DC pole \(|\lambda_j-1|<\delta_{\mathrm{DC}}\); select branch pole \(\widehat\lambda_\sigma\gets\arg\max_{j:\,|\lambda_j-1|\ge\delta_{\mathrm{DC}}} |a_j|\)
\EndFor
\If{\(\min_m|\widetilde{C}_\sigma(N_m)|<c_{\min}\) or \(\min(|\widehat\lambda_+|,|\widehat\lambda_-|)<\lambda_{\min}\)}
  \State Fall back to branch-linear readout [Eq.~\eqref{eq:vartheta_hat_branchlin}]
\EndIf
\State \(\widehat\vartheta_c\gets\Arg(\widehat\lambda_+\widehat\lambda_-^{*})/(2q)\) \Comment{\(\widehat\lambda_\sigma\) is the pole per sampling step \(q\), so the differential phase is divided by \(2q\)}
\State \textbf{Bootstrap:} for \(b=1,\ldots,B\), resample multinomial shot counts at each depth and target basis, rebuild \(\widetilde{C}_\sigma\) via Eq.~\eqref{eq:C_branch}, repeat the pole extraction; set \(\widehat\sigma_{\mathrm{boot}}=\mathrm{std}_b(\widehat\vartheta_c^{(b)})\)
\State \Return \(\widehat\vartheta_c\) (and \(\widehat\sigma_{\mathrm{boot}}\) if requested)
\end{algorithmic}
\end{figure}

\begin{figure}[t]
    \centering
    \includegraphics[width=\columnwidth]{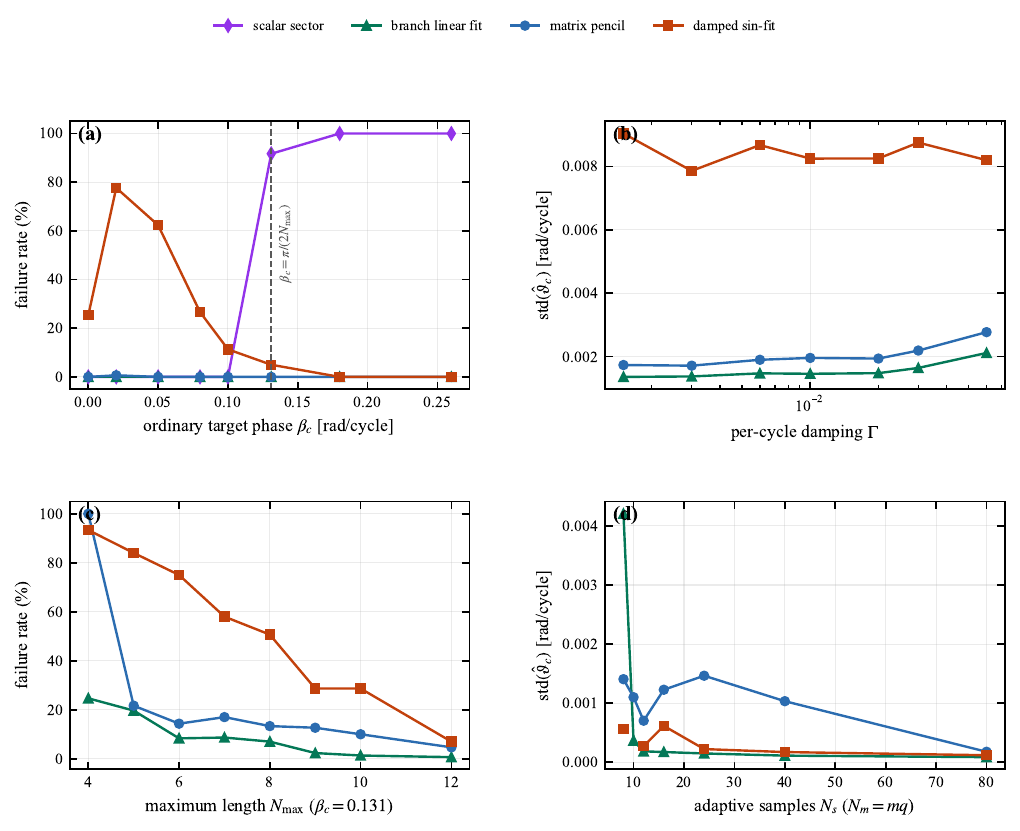}
    \caption{Estimator comparison under numerical stress tests
    ($\vartheta_c=0.03$~rad/cycle, 4000 shots, 300 independent trials).
    (a)~Failure rate versus ordinary target phase $\beta_c$.
    (b)~Standard deviation of $\varthetahat$ versus per-cycle damping
    $\Gamma$ at $\beta_c=0.05$.
    (c)~Failure rate versus maximum cycle length $N_{\max}$ at
    $\beta_c=0.131$.
    (d)~Standard deviation versus adaptive sample count $N_s$ with
    branch-asymmetric SPAM on $C_-$ ($=0.12$).
    Matrix-pencil and branch-linear readouts overlap within scatter;
    scalar-sector and sin-fit readouts are the dominant failure modes.}
    \label{fig:app_readout_comparison}
\end{figure}

Independent numerical stress tests compare scalar-sector, branch-linear, matrix-pencil, and damped sin-fit readouts applied to the same simulated branch sequences under four controlled degradations [Fig.~\ref{fig:app_readout_comparison}].
Each point aggregates many independent simulated trials over shot noise and SPAM realizations.
Branch-resolved fits agree within statistical scatter; scalar and sin-fit readouts fail under large target phase, damping, or sparse sampling.

\FloatBarrier
\section{Hardware platform and experimental details}
\label{app:hardware-experiment}

Hardware data were taken on QuantumCTek \hwcode{gd\_qc1} (\emph{Xiao Hong No.~1}; EZQ1.0; \hwcode{cqlib} v1.3.11) on 2026-07-07.
The device has 66 transmon qubits with tunable couplers; two-qubit gates are pulse-expanded \hwcode{CZ} blocks with offsets applied through \hwcode{GetPulse}.
Table~\ref{tab:hw-device} lists the experimental pair and baseline knob values at $\delta=0$.
Coupler G55 (Q36 control, Q30 target) was chosen from a 12-pair echoed-cycle survey (\minC\ $\approx 0.99$ on G55 versus $\lesssim 0.01$ on an initial G57 candidate).

\begin{table}[t]
\caption{Cloud processor and experimental pair (2026-07-07). Coherence, readout, and gate-error entries are device-level values reported by the backend resource panel on 2026-07-07; they are representative of \hwcode{gd\_qc1} and are not per-qubit calibrations of the specific Q30/Q36 pair.}\ 
\label{tab:hw-device}
\centering
\begin{ruledtabular}
\small
\begin{tabular}{@{}p{0.34\columnwidth}@{\hspace{0.35em}}p{0.58\columnwidth}@{}}
Item & Value \\
\hline
Machine & \hwcode{gd\_qc1} (66 qubits, 84 active couplers) \\
Coupler / qubits & G55: Q36 (control), Q30 (target) \\
Pulse index & \hwcode{qubit0}${=}$Q36 (control), \hwcode{qubit1}${=}$Q30 (target) \\
Readout & measure Q30 and Q36; bits decoded control-first (Q36, then Q30) \\
\hline
\multicolumn{2}{@{}l@{}}{\textit{Backend-reported representative values}} \\
Relaxation \(T_1\) & $27.96~\mu$s \\
Coherence \(T_2\) & $20.8~\mu$s \\
Readout error & $3.6\%$ (assignment fidelity $\approx 96.4\%$) \\
Single-qubit gate error & $0.11\%$ \\
Two-qubit gate error & $1.23\%$ (baseline \hwcode{CZ} fidelity $\approx 98.8\%$) \\
\hline
\multicolumn{2}{@{}l@{}}{\textit{CZ baseline at $\delta=0$}} \\
{\scriptsize\paramqzerodetune} (Q36) & $-0.035$ \\
{\scriptsize\paramqonedetune} (Q30) & $+0.131$ \\
{\scriptsize\paramczamp} & $-18.30$ \\
\end{tabular}
\end{ruledtabular}
\end{table}

Echoed cycles follow
\begin{equation}
  X_c X_t \;-\; \mathrm{CZ}_{\mathrm{pulse}}(\delta) \;-\;
  X_c X_t \;-\; \mathrm{CZ}_{\mathrm{pulse}}(\delta),
  \label{eq:echo-cycle-hw}
\end{equation}
with common offset \(\delta\) on both blocks [cf.\ Eq.~\eqref{eq:echo_cycle}].
Both qubits start in \(\ket{+}\); target \(X/Y\) pre-rotations and joint readout at \(N=1,\ldots,6\) supply the Pauli block (2048 shots per circuit).
Both qubits are measured (\hwcode{M Q30} and \hwcode{M Q36}); the emitted bit string is decoded control-first (\hwcode{control\_first}), i.e.\ the control Q36 bit precedes the target Q30 bit in the Pauli decoding, independent of the order in which the measurement instructions are listed.\ 
Here \(N\) counts echoed cycles [Eq.~\eqref{eq:echo-cycle-hw}], each comprising two native \hwcode{CZ} blocks, so the extracted \(\widehat\vartheta_c\) is the residual \(ZZ\) angle per echoed cycle; this is the quantity fed back and tabulated in Table~\ref{tab:closed-loop}, and it is not rescaled to a per-\hwcode{CZ} value.\ 
Branches \(C_\pm(N)\) are reconstructed offline via Eq.~\eqref{eq:C_branch} and \(\widehat\vartheta_c\) is extracted with the matrix-pencil readout of Algorithm~\ref{alg:pencil} (equivalent to branch-linear readout within Appendix~\ref{app:readout-comparison} scatter).

The four submitted stages were: (i)~actuator survey over five GetPulse knobs; (ii)~fine gain calibration of \paramqzerodetune; (iii)~injected-phase linearity at \(\vartheta_{\mathrm{inj}}\in\{0,\pm0.05,\pm0.1,\pm0.2\}\)~rad; (iv)~one native feedback step from the measured residual.\ 
Only \paramqzerodetune\ (the control-qubit Q36 detune, \hwcode{qubit0}) combined strong slope with preserved contrast (\minC\ $\approx 0.86$); \paramqonedetune\ (the target-qubit Q30 detune, \hwcode{qubit1}) collapsed the block (\minC\ $\approx 0.08$).\ 
This asymmetry is consistent with the normal form of Eq.~\eqref{eq:normal_form}: detuning the control qubit shifts its frequency relative to the coupler and thereby modulates the conditional \(ZZ\) interaction acquired during the \hwcode{CZ} pulse, moving \(\vartheta_c\) while the echo refocuses the accompanying local term; a target-qubit detune instead adds directly to the ordinary target phase \(\beta_c\), driving the \(\{IX,IY,ZX,ZY\}\) coherences out of the readout window faster than the echo can refocus and so collapsing the branch contrast.\ 
Fine calibration gave \(G_{\mathrm{inj}}\equiv|\partial\widehat\vartheta_c/\partial\delta_{\mathrm{detune}}|\approx 57.1\)~rad/(detune unit) with \(\partial\widehat\vartheta_c/\partial\delta_{\mathrm{detune}}<0\) (\(R^2\approx 0.998\)), and deliberate injection map \(\delta_{\mathrm{detune}}=-\vartheta_{\mathrm{inj}}/G_{\mathrm{inj}}\); logical \hwcode{CZ}/\hwcode{RZ} injection showed non-unit slope and was not used in Fig.~\ref{fig:guodun_hardware}.\ 
The correction applied in the closed loop is \(\delta_{\mathrm{detune}}=+\widehat\vartheta_c/G_{\mathrm{inj}}\) with no additional factor of two: the opposite sign to injection follows from the negative \(\widehat\vartheta_c\)--\(\delta_{\mathrm{detune}}\) slope, and both \(\widehat\vartheta_c\) and \(G_{\mathrm{inj}}\) are defined per echoed cycle [Eq.~\eqref{eq:echo-cycle-hw}], so the estimate, the gain, and the applied offset use the same per-echoed-cycle unit and compose consistently; the factor of two in \(\delta_{\mathrm{CP}}=-2\vartheta_c\) [Eq.~\eqref{eq:cp_vs_zz_angle}] relates \(\vartheta_c\) to the reported controlled-phase residual and does \emph{not} enter the actuator update.\ 

Table~\ref{tab:closed-loop} lists the three feedback rounds submitted after the native measurement, with the per-round bootstrap standard deviation of the matrix-pencil estimate.

\begin{table}[t]
\caption{Pulse feedback on G55. All \(\widehat\vartheta_c\) and bootstrap-std values are per echoed cycle [Eq.~\eqref{eq:echo-cycle-hw}, two native \hwcode{CZ} blocks] and quoted in radians.}\ 
\label{tab:closed-loop}
\centering
\begin{ruledtabular}
\small
\setlength{\tabcolsep}{3.5pt}
\begin{tabular}{@{}lcccc@{}}
Round & $\delta_{\mathrm{detune}}$ & $\widehat\vartheta_c$ & boot.\ std & \minC \\
\hline
0 (native) & 0 & $+0.0588$ & $0.16$ & 0.854 \\
1 (corrected) & $+0.00103$ & $+0.0036$ & $0.25$ & 0.865 \\
2 (over-corrected) & $+0.00109$ & $+0.0298$ & $0.16$ & 0.835 \\
\end{tabular}
\end{ruledtabular}
\end{table}

The one-step correction reduces the point estimate from \(\widehat\vartheta_c\approx+0.059\)~rad to \(\approx+0.004\)~rad, a point-estimate ratio of \(\sim\!16\times\) (full-precision value \(16.2\) from the unrounded round-0 and round-1 estimates, of which Table~\ref{tab:closed-loop} lists the rounded values); the bootstrap standard deviations (\(\sim0.16\)--\(0.25\)~rad) exceed both values, so this ratio is not statistically significant at the present shot budget and depth grid (see main text near Fig.~\ref{fig:guodun_hardware}).
Within this scatter Round~1 is already consistent with zero; the round-2 offset \(\delta_{\mathrm{detune}}\) is essentially unchanged from round~1 (\(0.00103\to0.00109\)~detune units), so under the deterministic actuator map its estimate should also remain near zero, and its larger point value \(+0.030\)~rad lies well within one bootstrap standard deviation of zero.
Round~2 is therefore a statistical fluctuation of an already-nulled residual, not evidence of a revised or failing actuator map; resolving any sub-\(0.06\)~rad structure beyond the round-0 suppression would require tighter statistics (more depths and repetitions) than a single closed-loop session provides.
We therefore present the closed loop as a sign-correct, one-step suppression demonstration and internal consistency check, not a high-precision residual measurement.
Raw summaries are archived as \hwfile{S6\_*\_G55\_*.json}.\ 

\end{document}